\documentclass[a4paper,11pt]{article}
\pdfoutput=1 

\usepackage{jcappub}

\usepackage[T1]{fontenc} 
\usepackage{graphicx}	
\usepackage{amsmath}	
\usepackage{amssymb}	
\usepackage{mathtools}
\usepackage{verbatim}
\usepackage{natbib}
\usepackage{color}
\usepackage{caption}
\usepackage{pdflscape}
\usepackage{subcaption}
\usepackage{rotating}
\usepackage{float}
\definecolor{darkblue}{rgb}{0,0.2,0.8}
\definecolor{darkred}{rgb}{0.75,0,0}
\def\simgt{\hbox{\,\rlap{\raise 0.425ex\hbox{$>$}}\lower 0.65ex\hbox{$\sim$}\,}}
\def\simlt{\hbox{\,\rlap{\raise 0.425ex\hbox{$<$}}\lower 0.65ex\hbox{$\sim$}\,}}

\title{\boldmath Dynamics of merging: Post-merger mixing and relaxation of an Illustris galaxy}

\author[1]{Anthony M. Young}
\author[1]{, Liliya L. R. Williams}
\author[2]{, and Jens Hjorth}

\affiliation[1]{School of Physics and Astronomy, University of Minnesota, 116 Church Street SE, Minneapolis, MN 55455, USA}
\affiliation[2]{Dark Cosmology Centre, Niels Bohr Institute, University of Copenhagen, Juliane Maries Vej 30, DK-2100 Copenhagen, Denmark}

\emailAdd{amyoung@astro.umn.edu}

\abstract{

During the merger of two galaxies, the resulting system undergoes violent relaxation
and seeks stable equilibrium. However, the details of this evolution are not fully
understood. Using Illustris simulation, we probe two physically related processes,
mixing and relaxation. Though the two are driven by the same dynamics---global
time-varying potential for the energy, and torques caused by asymmetries for angular
momentum---we measure them differently. We define mixing as the redistribution of
energy and angular momentum between particles of the two merging galaxies. We assess
the degree of mixing as the difference between the shapes of their energy distributions, $N(E)$s, and
their angular momentum distributions, $N(L^2)$s. We find that the difference is decreasing with time, indicating
mixing. To measure relaxation, we compare $N(E)$ of the newly merged system to
$N(E)$ of a theoretical prediction for relaxed collisionless systems, DARKexp, and
witness the system becoming more relaxed, in the sense that $N(E)$ approaches
DARKexp $N(E)$. Because the dynamics driving mixing and relaxation are the same, the
timescale is similar for both. We measure two sequential timescales: a rapid, 1 Gyr
phase after the initial merger, during which the difference in $N(E)$ of the two
merging halos decreases by $\sim 80$\%, followed by a slow phase, when the
difference decreases by $\sim 50$\% over $\sim 8.5$ Gyrs. This is a direct
measurement of the relaxation timescale. Our work also draws attention to the fact
that when a galaxy has reached Jeans equilibrium it may not yet have reached a fully
relaxed state given by DARKexp, in that it retains information about its past
history. This manifests itself most strongly in stars being centrally concentrated.
We argue that it is particularly difficult for stars, and other tightly bound
particles, to mix because they have less time to be influenced by the fluctuating
potential, even across multiple merger events.\\}

\begin{document}
\maketitle
\flushbottom

\section{Introduction}\label{Introduction}

Current theory states that galaxies formed in the potential wells of dark matter halos during hierarchical structure formation.  The structure and dynamics of these halos can 
provide understanding of the processes involved in galaxy formation and evolution as well as the nature of dark matter. Of particular interest are the 
central regions of halos where many competing and complementary processes between dark matter and baryons (and each type of matter with itself) shape galaxies.  Studies of galaxy 
formation and evolution are often performed with N-body simulations to capture the histories of structures like halos, that obviously 
cannot be directly followed over their billion-year timescales. Within these simulations, particles interact with their positions 
and velocities completely known.  This provides an excellent tool for investigating the time evolution of halo structure and dynamics.

N-body simulations provide important insights into dark matter halo structure.  These simulations show a near 
universal distribution of dark matter in halos, that is well described by phenomenological models over a few decades in radius \citep{Nav04,Sta09}.  Apart 
from halo structure, simulations provide a tool to investigate halo dynamics through the properties of the dark matter particles and their 
distribution functions \citep{Deh05, Vas08, Woj08, Nat97} as well as relationships between dynamics and structure \citep{Ton06}.  Simulations have also shown 
relationships between halo properties such as density, velocity dispersion, 
velocity distribution function, and velocity anisotropy \citep{Tay01, Han09, Mun13, Wil14}.  Additionally, several connections between galaxies and the halos 
they occupy have been probed through halo occupation modeling \citep{Kau97,Sel00, Hea16, Bos16, Fen16}.

Simulations have also been used to make predictions for direct detection of dark matter, which rely on the scattering of weakly interacting massive particles 
(WIMPs) \citep{Vog09}.  The halo dark matter velocity distribution and density are specifically applicable to the direct detection of dark matter on Earth.  
Experiments like CDMS \citep{Ake04}, PICO \citep{Amo16}, CRESST \citep{Ang02}, and their successors need to model the total flux at Earth of WIMPs to predict 
what signals may look like.  The recoil spectrum is modeled as a function of two dark matter properties, the halo density and the velocity distribution.   
The halo density can be inferred from observations and simulations, but the velocity distribution has little to no observational basis and is generally 
determined through simulations \citep{Lew96}. 

As computational power has improved, allowing for marked increase in simulation complexity and resolution, additional physics has been added to these simulations to try to model nature more accurately.
Specifically, baryons and their physics have been introduced to investigate the co-evolution of dark matter and baryonic matter 
structures \citep{Gne04,Ped09,Duf10,Bro12,Saw13,Cus14,DiC14,Vel14,Saw15,Sch15a}.  Baryons are important for several reasons, as baryonic processes can alter 
the distribution of matter, especially in the central regions of halos \citep{Nav96,Gne02,Rea05,Mas08,Pon12,Tey13,Nip15,Ono15,Rea16}.  Baryons can condense 
through radiative cooling and conversely, their density can decrease through feedback from active galactic nuclei or 
star formation.  The central regions of halos have unique physics and properties \citep{Xu17} because of these various processes.   It is these structures 
that we will probe over a galaxy's lifetime.

Central baryons often produce a change in the density slope 
of these systems, that marks the transition from the baryon dominated central region to the rest of the halo (see Figure \ref{density}).  This feature, 
which we call an `oscillation', should be erased as the halo becomes completely relaxed.  A recent compilation of models for relaxed systems presented 
in Beraldo e Silva et al. \citep{Ber13} show no such features present.  This implies if the systems we study have such features, they must 
be erased by the time the halo becomes fully relaxed.  The persistence of this baryon/dark matter transition implies that the relaxation in the central parts is incomplete.  As 
in Lynden-Bell \citep{Lyn67}, we assume the process of violent, or collisionless relaxation 
is driven by particles exchanging energy with the time-varying global potential.  The final configuration produced by this violent relaxation, if given enough time, will be a fully 
relaxed system.  A fully relaxed system should not contain any information about its past history and assembly, and its particles should not exchange 
energy or angular momentum.  We consider a system relaxed if its energy distribution follows that of DARKexp (see below).  Why halos do not fully relax is 
one question we would like to investigate with the analysis below.  Though not a topic of this paper, we note that similar oscillation-like features in the 
density profile slope are present in pure dark matter halos, where they also point to departures from a fully relaxed state \citep{You16}.

One of our main 
goals will be to understand the departures of a halo from a fully relaxed state, and use these departures to provide insight into the dynamical state of the 
system during and after a merger.  Previous work has proposed ideas related to incomplete relaxation, including distribution function and energy distribution 
features related to assembly history \citep{Vog09}, and incomplete relaxation based on energy \citep{Hjo91}.

Related to relaxation, is the process of particle mixing, which we will also track in subsequent sections.  The mixing we will discuss is 
different from the phase space mixing presented in Tremaine et al. \citep{Tre86}, as ours involves the mixing in energy and angular momentum between particles in two merging 
halos.  We say that the particles of two merging halos are mixing in energy if the two sets of particles' energy distributions are approaching the same shape 
following the merger.  The dynamics causing mixing and relaxation are the same; the difference in how we measure the two is described in Section 2.

Mergers are an important driving force shaping halos as they grow.  Halo mergers are 
usually described as minor or major depending on the mass ratio of the merging halos.  Mergers with near equal mass halos, $1\lesssim M_1/M_2\lesssim3$, are 
described as major, and those with a dominant mass halo absorbing a smaller halo are labeled as minor.
Of particular interest within halo evolution are the effects mergers have on a halo's final configuration.  For example, massive elliptical galaxies around redshift $z=2$ 
have been found to be more compact than similar mass ellipticals in the local universe \citep{Tru07, Bui08, Cim08, van08,Tof14}. The disconnect between the abundance of high redshift 
compact ellipticals and the lack of compact ellipticals in the local universe may indicate a growth with time, or ``puffing'' up, as compact ellipticals 
evolve to their present day form through minor mergers and continual accretion \citep{van08,Naa09, Bar13}.  The formation and evolution 
of these compact ellipticals are also not completely understood.  Recent efforts have used simulations to inspect their formation and found gas 
rich major mergers between $2<z<4$ as one of two proposed explanations, along with early formation time \citep{Wel15}. 

The simulation we will use to investigate merger-driven halo evolution is Illustris.  Illustris is a suite of hydrodynamical N-body 
simulations of galaxy formation \citep{Vog15}.  The specific simulation we will analyze in later sections, Illustris-1, contains dark matter particles as well as baryons 
in the form of stars and gas.  The simulation also tracks supermassive black holes, but they are not considered in this analysis 
because their mass is a small fraction of the halo central region's mass ($\sim10^{-4}M_{200}$ for halos at $z=0$).  Illustris-1 is the highest resolution of the Illustris simulations and contains 
$2\times1820^3$ total particles in a $(106.5\text{Mpc})^3$ volume \citep{Vog14}.  As the simulation evolves, subhalos and halos are identified with a 
friend-of-friend (FoF) algorithm \citep{Dav85} and gravitationally bound substructures are found using the SUBFIND algorithm \citep{Spr01,Dol09}.  Merger 
trees are calculated using a SubLink algorithm and provide merger histories for subhalos \citep{Rod15}.  
Our goal is to describe the merger process in Illustris halos and galaxies by looking at the evolution of energy and angular momentum distributions with a focus on 
post-merger relaxation.  We also investigate changes in the structure of the halo to further explore the connection between dynamics and structure.  

What distinguishes our work from others on similar topics is that we judge the relaxation state by comparing simulation results to the theoretically derived 
model for the structure of self-gravitating isotropic collisionless (infinite N) systems, called DARKexp \citep{Hjo10,Wil10}.  DARKexp, which is based on maximum 
entropy and follows similar statistical mechanics arguments as the derivation of Maxwell-Boltzmann distribution, provides excellent fits to 
simulated and observed dark matter halo energy distributions and density profiles \citep{Ber13,Hjo15,Ume16,Nol16,You16}.  Its energy distribution is given by
\begin{equation}
 N(\epsilon)\propto e^{\phi_0-\epsilon}-1
\end{equation}
where $\phi_0$ and $\epsilon$ are the halo dimensionless central potential and energy, respectively.  It is important to note that DARKexp is not a phenomenological model, and that it has only one shape parameter, $\phi_0$.  

Because the prediction is in terms of energy ($E$), we use energy and 
angular momentum squared ($L^2$) in our analysis, and not phase-space or configuration space parameters.  As all energy distributions, DARKexp $N(E)$ is insensitive to 
velocity anisotropy, meaning that while derived from isotropic assumptions, it should also describe anisotropic halos.  This will 
allow us to fit DARKexp $N(E)$ to Illustris halo energy distributions.  Previous work has extended DARKexp $N(E)$ to include angular 
momentum in the distribution, to provide a more complete dynamical description of systems \citep{Wil10b,Wil14}
and we further investigate the halo $N(E,L^2)$ in our analysis below. 

This paper studies the redistribution of particles' $E$ and $L^2$ during the merger process to understand the halo's approach to a relaxed state after a 
merger.  We would like to address (i) how does this happen, and (ii) how fast does it happen.  In Section 2, we show
how dark matter particle energy and angular momentum change by studying their redistribution during the merger history of an Illustris halo.  We then devise 
metrics to quantify mixing and relaxation of the dark matter halo, and address how merging dark matter particles deepen in the potential well as they are absorbed by 
the main halo.  Section 3 focuses on the role of baryons during relaxation.  We then 
quantify how the central region of the halo evolves in the context of the halo as a whole.  Our conclusions and additional discussions are presented in 
Section 4.

\section{Dynamical evolution of dark matter particles in merging halos}

We are interested in the changes in a halo's structural and dynamic state as it nears equilibrium and moves toward a relaxed state.  We approach this 
problem by finding a halo whose history contains isolated, simple merger events where only a few distinct smaller halos merge with a central halo.  
A simple merger history will help with the analysis and interpretation below.

\begin{figure}
\centering
\includegraphics[scale=.5]{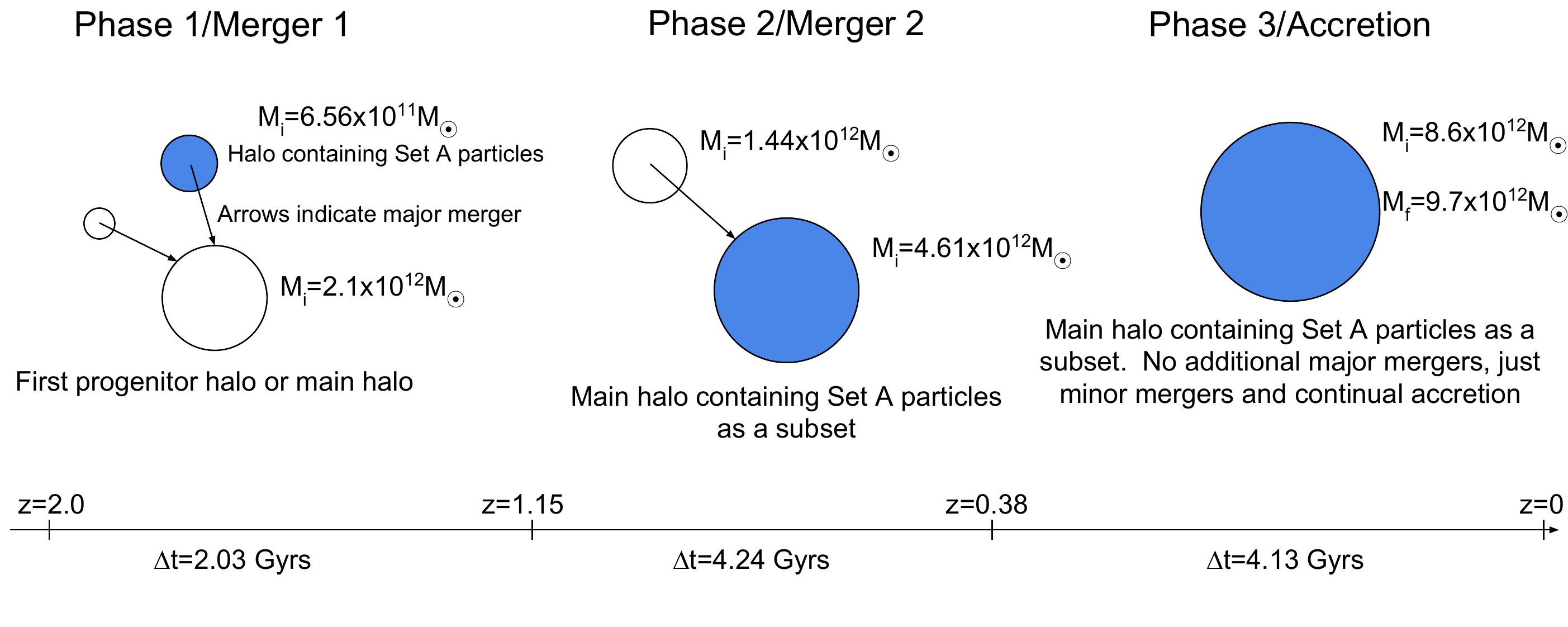}
\caption{Three physically motivated phases that capture different major merger events in halo 138's merger history.  The blue shaded circle is the halo that 
contains Set A particles during that phase, and arrows indicate major mergers with the main halo.  The initial halo total mass in that phase, $\text{M}_i$, is given for each phase
and $\text{M}_f$ is the final halo total mass during the Accretion phase.  The redshifts mark the boundaries between the phases and the phase duration is given in Gyrs.}
\label{mergerh}
\end{figure}

In our search we found halo 138.  It has a quiescent merger history compared to other halos of similar mass; it underwent two distinct major merger 
events with the last major merger happening several Gyrs ago.  We acquired its merger history using the SubLink trees \citep{Rod15} in Illustris and split 
its evolution history into three physically motivated phases.  A schematic of the 3 phases is presented in Figure \ref{mergerh}.  The arrows indicate 
major mergers, with the first merger event in Phase 1 and the second event in Phase 2.  The blue shaded circles are the halos containing the Set A particles 
in their given phase.  Set A particles will be defined in the following paragraph.  

Phase 1, which we will also call Merger 1,  corresponds to a merger event that starts around $z\sim2.5$ with two smaller halos merging with the first progenitor halo in the main progenitor 
branch of halo 138.  The definition of the first progenitor halo is described in De Lucia \& Blaizot \citep{DeL07} and identifies the progenitor halo of the halo in question (in our 
case halo 138) that has the `most massive history'. The most massive history is the branch of the merger tree with most of the mass of the final system and 
this branch is called the main progenitor branch. The first progenitor halo is defined at each redshift.  We will refer to the first progenitor halo as the 
main halo since other halos are merging with it. We use the dark matter particles of one of the two merging halos, whose mass is a factor of 3.2 smaller than 
that of the main halo, for much of our analysis. We will refer to these approximately $1.2\times10^5$ particles as Set A particles. 

\begin{figure}
\centering
\includegraphics[scale=.4]{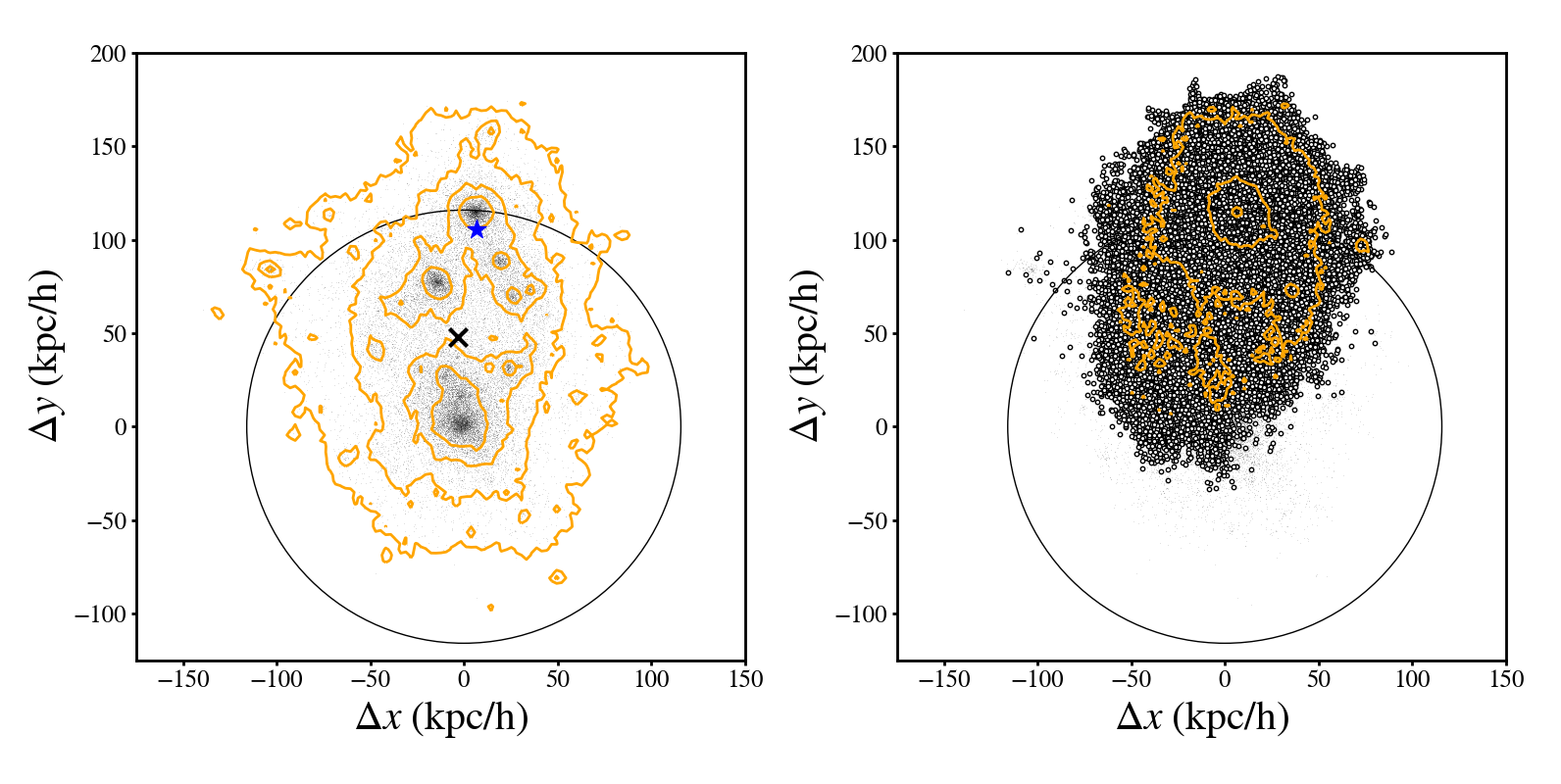}
\caption{Left: Dark matter density projected onto the $x$ and $y$ axis where $(x,y)=(0,0)$ is the center of the potential well of halo 138 
but at $z=2.0$.  The black x marks the center of mass for the entire halo and the blue $\star$ marks the center of mass for the merging halo that contains the Set 
A particles.  The black circle indicates the halo virial radius.  Right: The same 
physical extent except the points are the dark matter particles of the merging halo (Set A particles) with its center located at around 
$(x,y)\sim(10,110)$ (in kpc/$h$) on the virial radius circle of the halo.  The points are shown to indicate the spatial extent of the merging 
particles as the center of mass of the merging halo moves closer to the main halo center.  The contours indicate regions of high number density of the 
points.  Judging from the extent of the particles, tidal stripping is present.}
\label{potential68}
\end{figure}

While the merger in Merger 1 starts prior to $z=2$, as the two halos are moving towards a merger, we use a different starting event.  We 
define the end of the pre-merger phase, which we will use as the start of Merger 1, as the instance where the merging halo center of mass crosses within $r_{vir}$ of the 
main halo center, where $r_{vir}$ is the radius that encloses 200 times the critical density at that redshift.  We can visually identify the pre-merger step by looking at a plot of the 
density of the halo at $z=2.00$, projected in the simulation $x-y$ plane.  The left panel in Figure \ref{potential68} shows the projected 2D density of the system 
along with a black `x' to denote the center of mass for the whole halo, and a blue `$\star$' to denote the center of mass of the merging halo (Set A particles).  
The right panel is the same as the left, except the points show a spatial extent of the Set A particles contained in the merging halo that are 
now a part of the larger, combined halo, along with contours to show their density. 

Merger 1 ends at $z=1.15$ when the centers of mass of the merging halo containing Set A particles and main halo are sufficiently close, which we 
define as within $0.07r_{vir}$.  This definition is similar to the halo relaxation criterion used in Neto et al. \citep{Net07}, 
where a halo is considered relaxed if $s<0.07$, where $s=|\boldsymbol{r}_c-\boldsymbol{r}_{cm}|/r_{vir}$, and describes a normalized offset between the center of mass 
and the location of the deepest potential of a halo.  We call the period after this criterion is met the post-merger phase.
We emphasize that the merging halo particles will still continue to relax and move in the $E-L^2$ space beyond this point in time.  

Just as the first merger reaches completion, the second merger starts at redshift $z=1.07$.  This marks the beginning of Phase 2, or Merger 2.
The second merger event sees the product of the first merger, now merging with another halo.  Phase 3, or Accretion phase, begins when the second major merger is complete at $z=0.38$ and ends at the present.  This phase sees no major mergers, only the gradual 
movement towards relaxation, although minor mergers and accretion still occur.  This type of accretion should be well described by the virial theory 
argument presented in Naab et al. \citep{Naa09}, and meant to explain the puffing up of compact, $z=2$ galaxies.  


The fractional mass increase, $\eta$, during Accretion phase is 9.12\%.  If the ratio of the mean squared speed of 
the accreted material to that of the initial material is small, the ratio of initial and final radii is $r_f/r_i=(1+\eta)^2$.  This would produce a 
final $r_{vir}=333$kpc/$h$.  This is close to the actual growth, as $r_{vir}$ grew by 16.36\% from around 280 kpc/$h$ to 326 kpc/$h$.

The specific start and stop times of each phase can be seen in Figure \ref{mergerh} along with some additional information. 


We concentrate on answering two questions in this Section: (i) how do the energy and angular momentum of dark matter particles evolve in response to 
mergers and quiescent accretion, and (ii) what is the time scale for the relaxation process.  We do this by tracking the energies and angular momentum 
squared of Set A particles during the 3 phases defined above.

\subsection{Evolution in $E-L^2$ space}\label{EL2evo}

First we need to calculate $E$ and $L^2$ for all halo particles, relative to the halo center, which is taken to be the halo's deepest potential.  All energies 
and angular momenta are actually specific energies and specific angular momenta, but denoted from now on as $E$ and $L^2$.
From Illustris \citep{Nel15,Vog15}, we took particle positions, velocities, and potential energies.  Kinetic energy is calculated from particle 
velocities, corrected for the halo central bulk motion.  The bulk motion is taken as the average velocity of particles within 
10\% of the halo virial radius, $r_{vir}$.  Illustris does 
provide a mass weighted average halo velocity, but those values and our $r<0.1r_{vir}$ weighted values can sometimes differ depending on the dynamics of the 
central subhalo compared to the halo as a whole.  They tend to agree later in time as the halo is more relaxed.  We use these central region weighted 
velocities because we are most interested in the particles in this region.  We use the same corrected velocity to 
calculate particle angular momentum.  We did not consider angular momenta before the two halos met our pre-merger criterion. 

\begin{landscape}
\begin{figure}
\vskip -2.4cm

\begin{subfigure}{.75\textwidth}
\includegraphics[scale=.27]{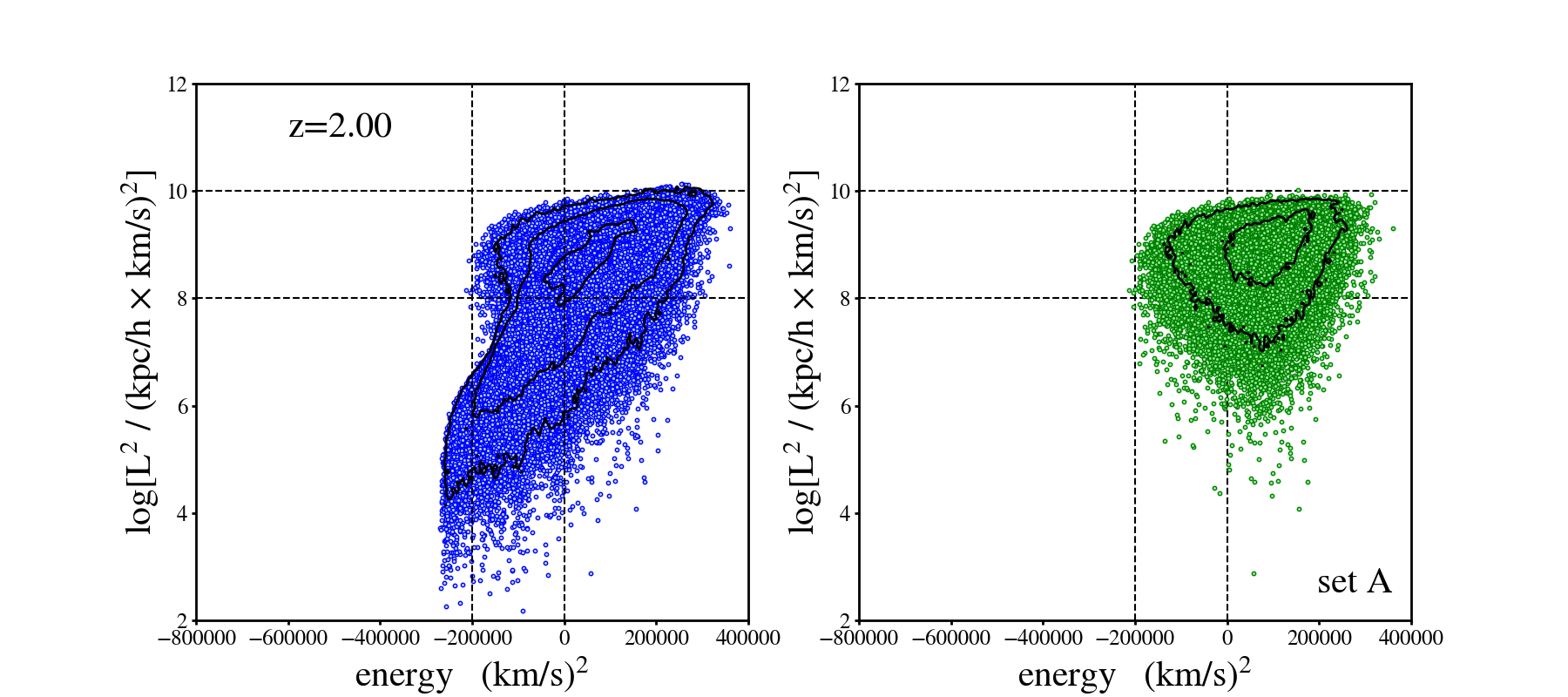}
\includegraphics[scale=.27]{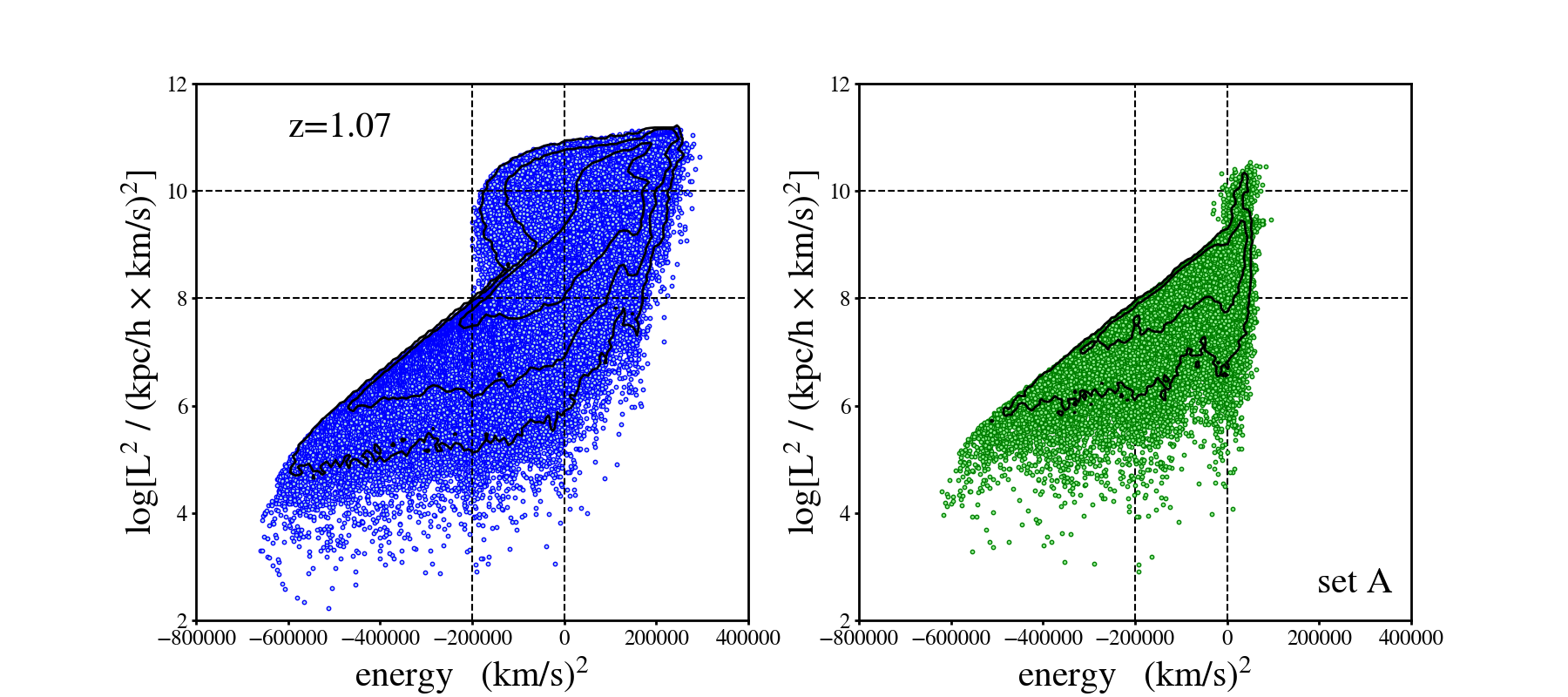}
  \includegraphics[scale=.27]{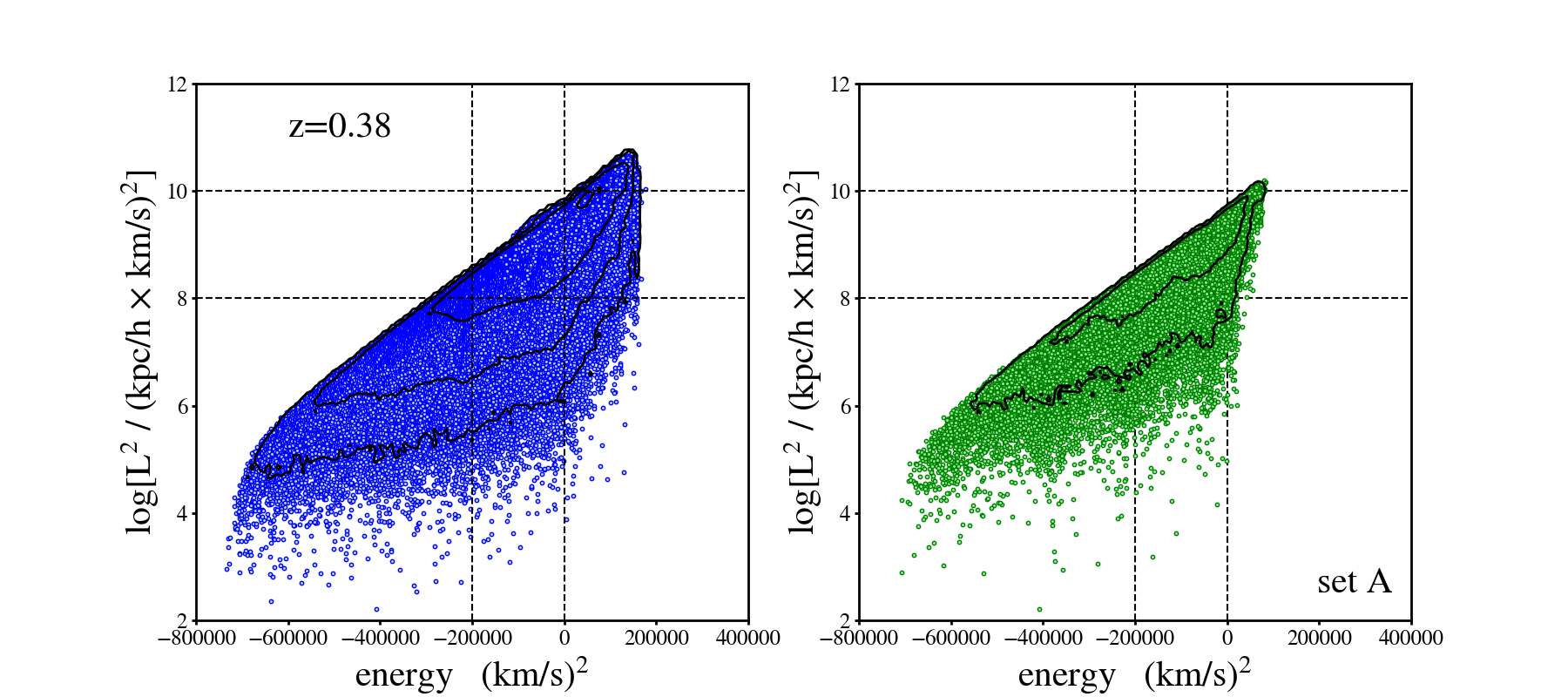}

  \label{fig:sub1}
\end{subfigure}%
\begin{subfigure}{.25\textwidth}
\includegraphics[scale=.27]{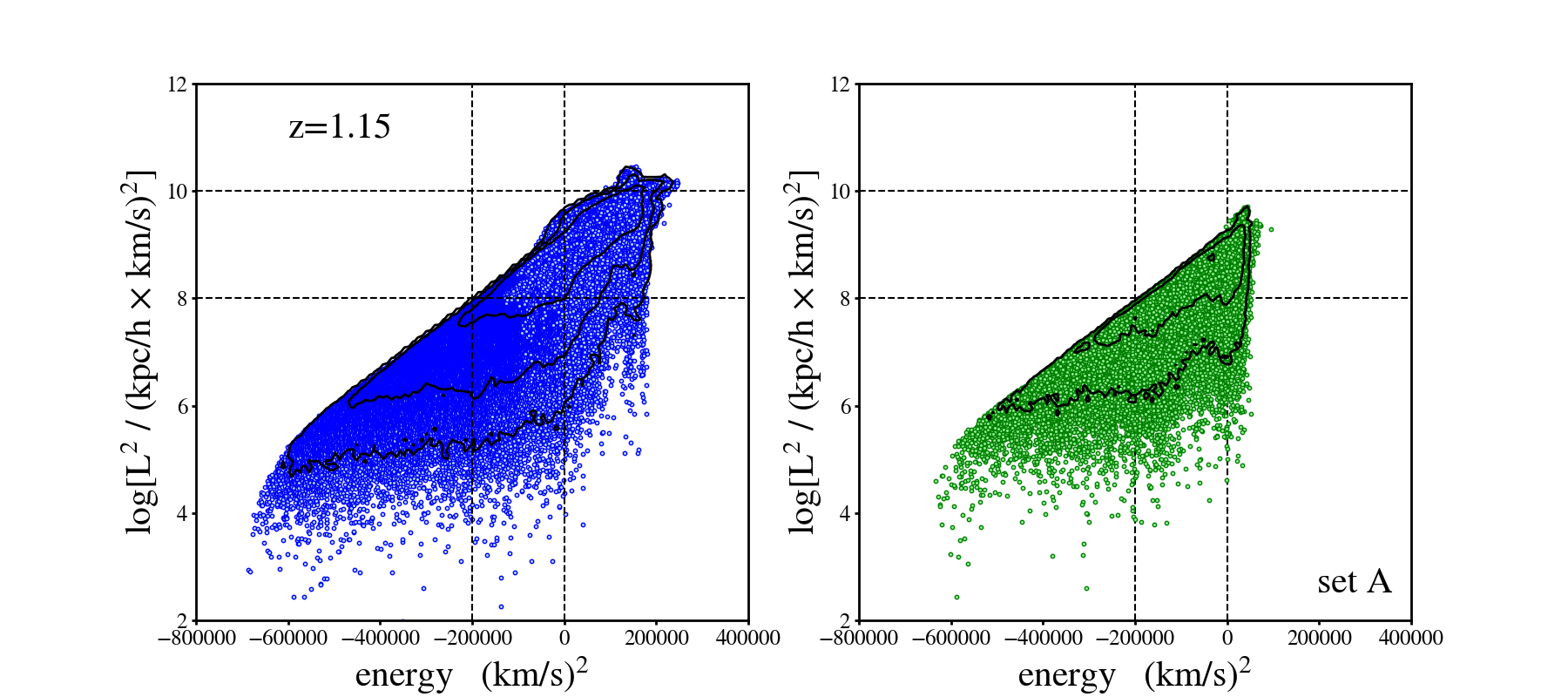}
  \includegraphics[scale=.27]{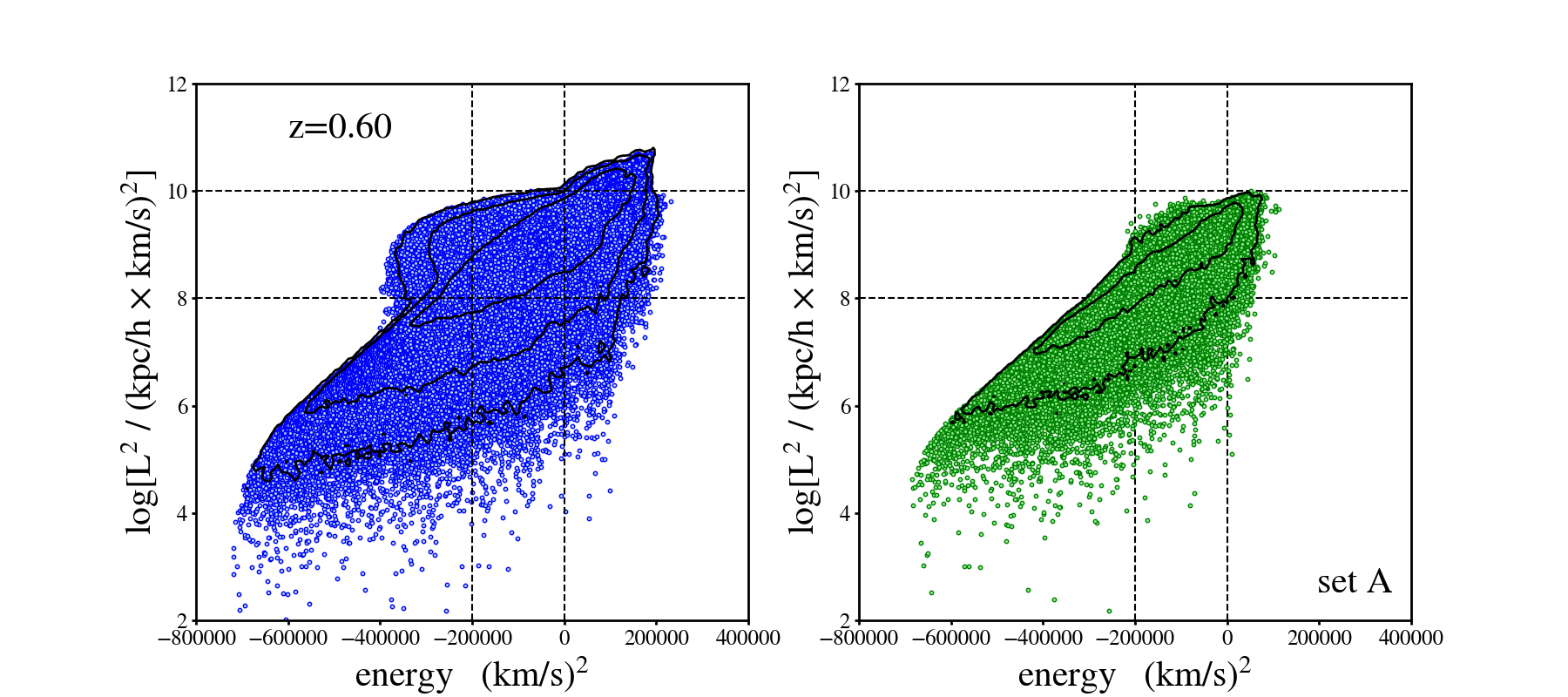}
  \includegraphics[scale=.27]{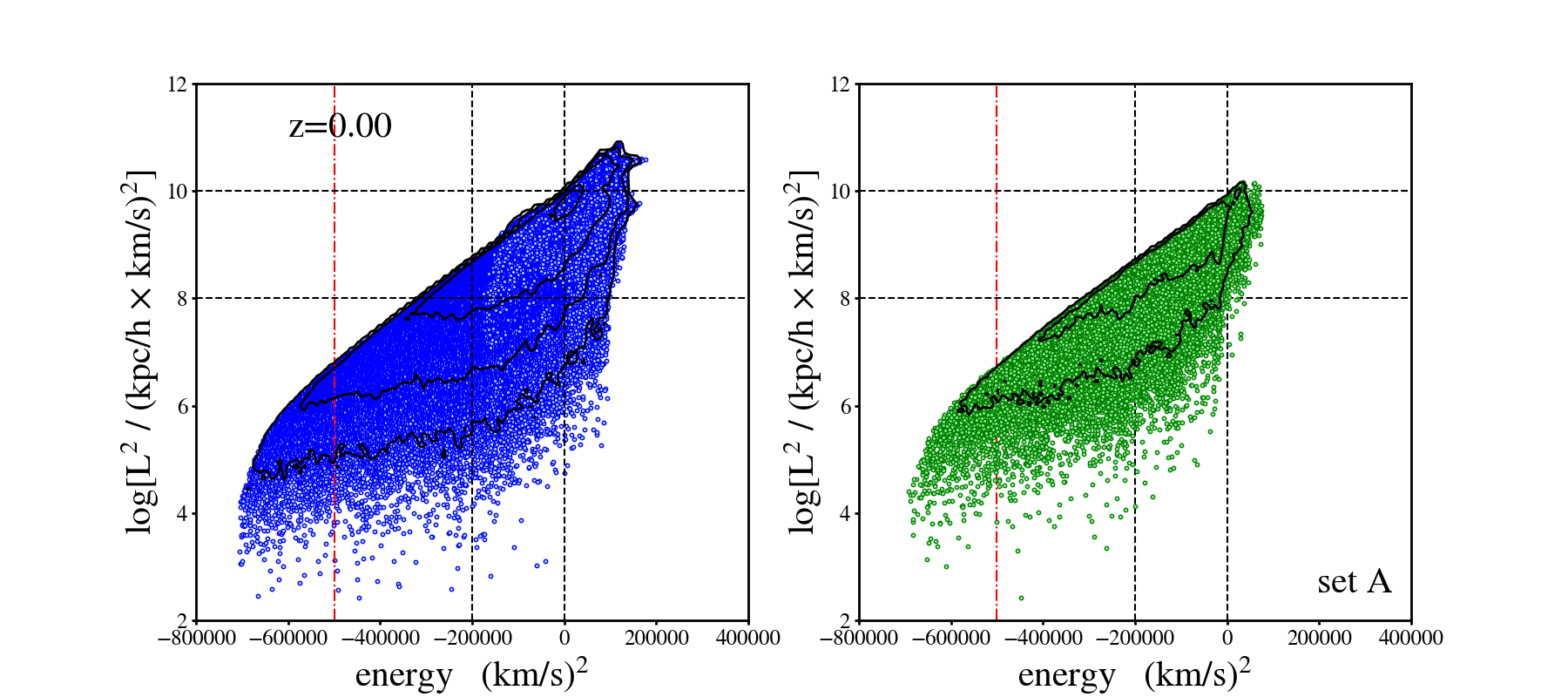}

  \label{fig:sub2}
\end{subfigure}
\hskip -1.0cm
\caption{$E-L^2$ plots for all halo dark matter particles in blue (left of each set of 2 panels) and Set A particles in green (right) to show 
their extent in that space for several redshifts.  Contours show regions of high number density.  The 
black vertical and horizontal dashed lines are guidelines to help show the bulk motion of the points between redshifts.  During earlier 
epochs, one can see the bump of the merging halo in the blue panels,  when compared to the $z=0$ distribution.  This bump moves in $E$ and $L^2$ and dampens out over time as the merging 
halo completes its merger with the main halo.  The vertical red dash-dot line in the $z=0$ panels at $E=-5\times10^5$(km/s)$^2$ indicates a region to 
the left of the line where particles are well mixed in $L^2$; (see Section \ref{DML2relax}.)}
\label{L2vEcontour}
\end{figure}
\end{landscape}

Figure \ref{L2vEcontour} shows $E$ vs $L^2$ for all halo 138 dark matter particles at redshifts corresponding to the start and end epochs of the 3 Phases, 
along with one intermediate redshift $z=0.60$.  Each of the sets of two 
panels shows all dark matter in blue (left) and Set A particles in green (right).  The contours show the highest number density regions in their respective panels.
The dashed black guidelines in all the panels have the same values and serve to emphasize the bulk motion of all particles in the space.  
Over time, the halo continually gained mass and deepened its potential well.  As a result, across all phases, the general trend is a move in the median energy 
value to more negative, bound energies.  

Angular momentum, on the other hand, shows more complex behavior.  During Merger 1, Set A particles tend to lose angular momentum during their merger with the 
main halo 
\noindent
and correspondingly, their median $L^2$ value decreases.  After this merger, the median $L^2$ 
value of Set A particles increases in Merger 2 as newly merging particles lose $L^2$ to the main halo, which now includes Set A particles, just as Set A particles did in Merger 1.  
Similarly, the median $L^2$ value for the main halo particles tends to increase over all time, presumably at the expense of merging particles as they 
lose $L^2$ during their infall.  We will call the main halo dark matter particles that are not Set A particles, Set A$^c$.  Note that Set 
A$^c$ changes its membership slightly as some particles enter the main halo through smooth accretion, while others get ejected.

Examining the $E-L^2$ distribution of halo 138 early during Phase 
1 reveals that Set A particles are quite localized in that space (Figure \ref{L2vEcontour}).  As Set A particles merge with the main halo, their most bound dark matter particles 
stay gravitationally bound together until around $z\sim1.36$. The transition from infall to the break up and assimilation of 
the core can be seen in the shape of the halo's $N(E,L^2)$ distribution, as it changes from its initial configuration at $z=2.00$, to eventually appear 
like that of the total dark matter population at $z=1.15$.  During this merger, the total halo dark matter $N(E,L^2)$ is a 
superposition of two individual halos' $N(E,L^2)$ with the Set A particles appearing as a bump on the $L^2$ envelope (at large $L^2$ values) of the main 
halo.  For the first major merger, this feature dampens over time and eventually completely disappears, with the final configuration seen in the $z=1.15$ 
panel.  A similar disappearance occurs with the second major merger, where the merging halo is not noticeable anymore by $z=0.38$. 

\begin{figure}
\centering
\includegraphics[scale=.4]{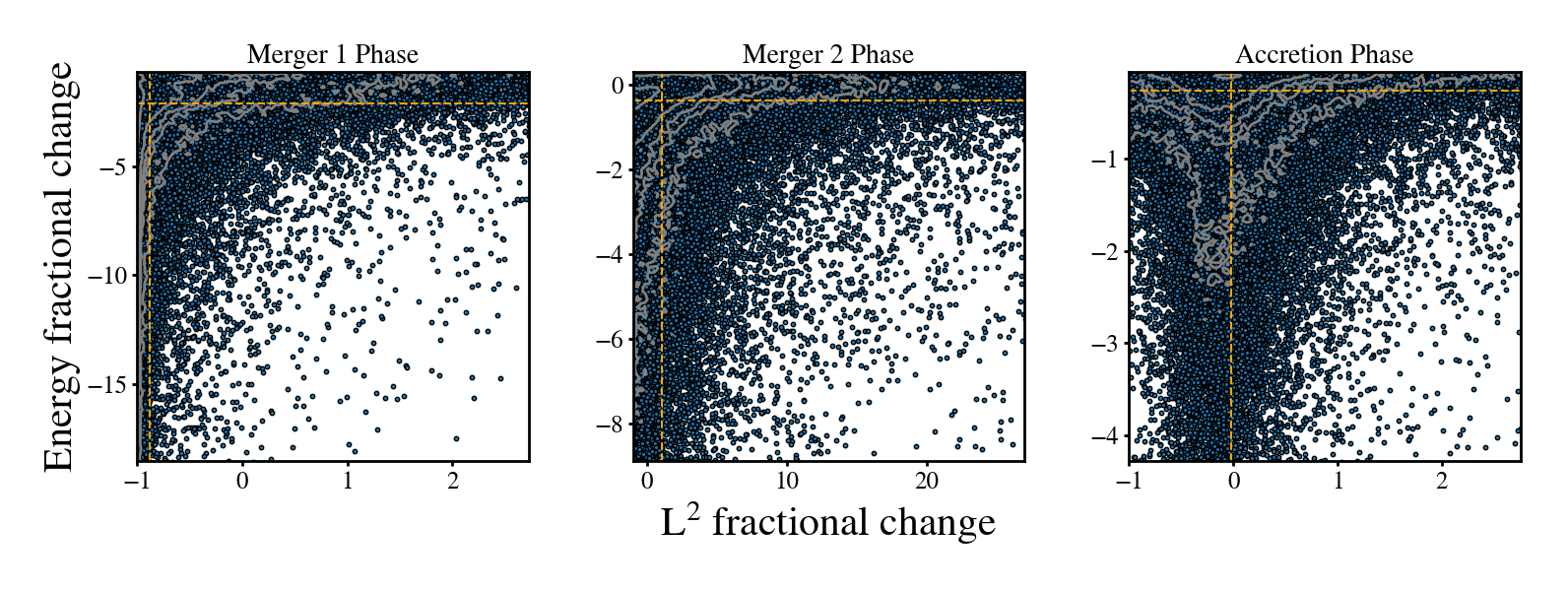}
\caption{The fractional change in $E$ and $L^2$ of the Set A particles during Merger 1 (left), Merger 2 (middle), and Accretion phase (right), as described in Figure \ref{mergerh}.  The 
extent of the plot window shows the 5\% to 95\% range of particles in each axis.  The contours show regions of higher point density.  The orange dashed lines 
indicate the median values for the fractional change in energy: $-2.11$, $-0.38$, and $-0.26$, and the fractional change in $L^2$: $-0.88$, 1.10, $-0.02$ for Merger 1, 
Merger 2, and Accretion phase, respectively.  These values are not per unit time; see Figure \ref{mergerh} for Phase durations.  Note that the median $f_E$ changes monotonically with time, whereas $f_{L^2}$ does not.}
\label{dEL2}
\end{figure}

Another way of looking at the motion of halo particles in the $E-L^2$ space over time is to track the fractional change in these quantities.  The 
fractional change is calculated as the difference between the initial and final quantity divided by the absolute value of the initial quantity. For example, 
the fractional change in energy is given as
\begin{equation}
 f_E=\dfrac{E_f-E_i}{|E_i|}
\end{equation}
with the same definition used for the fractional change in $L^2$ giving $f_{L^2}$.  This allows us to analyze how $L^2$ and $E$ are redistributed on a particle basis. For initial 
and final redshifts, we use the phases 
outlined in Figure \ref{mergerh}. For Merger 1, we use the Set A particles to probe a merger event and follow their infall towards the main halo.  In Merger 2, since the 
Set A particles are already part of the newly formed halo, we can track how they are disturbed when a new major merger occurs.  Finally in Accretion phase, after the halo 
has gone through its last major merger, we can examine how these particles move towards a relaxed state.  Fractional changes $f_E$ and $f_{L^2}$ for 
the three phases can be seen in Figure \ref{dEL2}.  The panels show each Set A dark matter particle 
as a blue point, with contours indicating regions of higher density. The orange dashed lines show the median values of the particles' $E$ and $L^2$.  

As we already saw earlier in this Section and in Figure \ref{L2vEcontour}, the majority of the Set A particles lose energy and $L^2$ as they move towards 
the larger main halo (Figure \ref{dEL2}, left panel).  Any asymmetry in the system will lead to torques and angular momentum transfer, but during a merger 
one of the main mechanisms to transfer angular momentum is dynamical friction \citep{Cha43,Bin87}.  In Merger 1, the median value for $f_E$ is $-2.1$ and 
for $f_{L^2}$ it is $-0.88$.  

The beginning of Merger 2 corresponds to the point in time when Set A particles have merged with the original main 
halo to form a new halo. At this time ($z=1.15$), another halo begins its merger. As it moves towards the main halo, the main 
halo particles are disturbed.  This is noticeable not only in a density map of the halo, but in the $E-L^2$ space as well (middle panel of Figure \ref{dEL2}).  The 2D distribution looks similar to the Merger 1 
distribution with two main exceptions.  First, there is now a small population of particles that gain energy ($f_E>0$), although the median value for 
$f_E$ in Merger 2 is $-0.38$.  This is caused by the new infalling particles exchanging 
energy with the more negative energy main halo particles.  Second, the infalling particles also exchange angular momentum with the main halo particles.  This 
exchange was observed in Merger 1, only the Set A particles were giving up $L^2$ since they were merging.  Now they are receiving $L^2$ from the new merging 
particles.  As a result the median $f_{L^2}$ value of Set A particles is now positive (1.1), whereas it was negative in Merger 1. 

Once Merger 2 has ended, in the Accretion phase, the halo begins its steady approach towards equilibrium and relaxation.  While the halo is still accreting additional mass, no 
new major mergers occur. This phase captures how particles redistribute in response to accretion and minor mergers.  The right panel of Figure \ref{dEL2} shows that the majority of our Set A particles still lose energy, but are evenly split between 
gaining and losing $L^2$. The negative values for $f_E$ imply that the halo is still accreting matter throughout this phase,
thus deepening the relative potential.  The $f_{L^2}$ distribution, while not exactly centered on $f_{L^2}=0$, is now more symmetric between those 
gaining and losing $L^2$, indicating that the halo is becoming more mixed in angular momentum.  However, since the halo is not spherically symmetric, we cannot 
disentangle the global redistribution following a merger from the fact that $L^2$ for a given particle in a triaxial system will change along its orbit.

Over the whole evolution, the median values for $f_E$; $-2.11$, $-0.38$, and $-0.26$, and $f_{L^2}$; $-0.88$, 1.10, $-0.02$ show dissimilar behavior.  The energy, while always 
decreasing, is doing so by smaller amounts each subsequent phase.  Interestingly, both major mergers seem to have no effect on the sign of the median particle's change in energy, 
whereas the median value of angular momentum is changing signs.  Median change in angular momentum depends on whether the particle is merging with the main halo or already a part 
of the main halo.

\subsection{Mixing of halo dark matter particles}\label{DMmix}

During Accretion phase, we observed evidence of the halo becoming more mixed in angular momentum, because $f_{L^2}$ was evenly split between positive and negative values. 
This supports the idea of a halo moving toward relaxation.  Now we want to generalize this analysis. To do this we use the halo $N(E,L^2)$ to diagnose the 
system's state as a function of time, by comparing Set A and Set A$^c$ of the halo dark 
matter particle population.  We assume that if they are well represented by the same distribution, they are well mixed.  While the mixing of particles in $E$ and $L^2$ is 
not sufficient to describe a system as relaxed, it is a necessary condition for relaxation.  

\subsubsection{Mixing of particle angular momentum}\label{DML2relax} 

The right panel of Figure \ref{dEL2} suggests that during Accretion phase at least some subset of particles are fully mixed in $L^2$.  To test that further, we use a 
non-parametric Kolmogorov-Smirnov test (KS test) to assess the likelihood that Set A and Set A$^c$ come from the same distribution.  We divided $N(E,L^2)$ 
distributions of 
Set A and Set A$^c$ particles into 80 energy bins of constant $\Delta E$ to define $N(L^2)$ for a given energy.  Each bin has tens to hundreds of particles at the most 
bound energies, and $1\times10^4$ to several $10^5$ particles in the less bound energy bins for the $z=0$ halo.  We applied a KS test to the two 
$N(L^2)$ distributions within each energy bin and found that the null hypothesis, that both sets are drawn from the same parent distribution, cannot be 
rejected for particles more bound than $\approx2/3$ of the halo's most bound energy, at $z=0$.  This result was consistent when we changed 
the number of bins for the KS test and when we applied the Anderson-Darling test as well.  The region to the left of the red dash-dot 
vertical line in the $z=0$ panel on Figure \ref{L2vEcontour} indicates where the halo is well mixed in $L^2$ for a given energy.  These particles occupy the 
densest region of the halo.

\subsubsection{Evolution of dark matter energy distribution shapes}


\begin{figure*}
\centering
\begin{subfigure}{.5\textwidth}
  \centering
\includegraphics[width=\columnwidth]{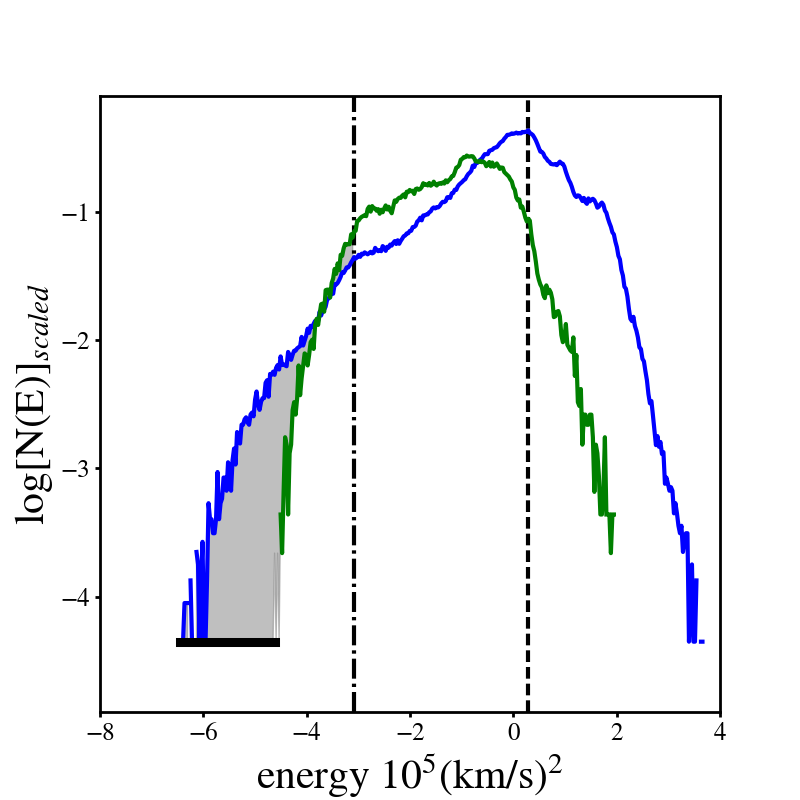}
  \label{fig:sub1}
\end{subfigure}%
\begin{subfigure}{.5\textwidth}
  \centering
\includegraphics[width=\columnwidth]{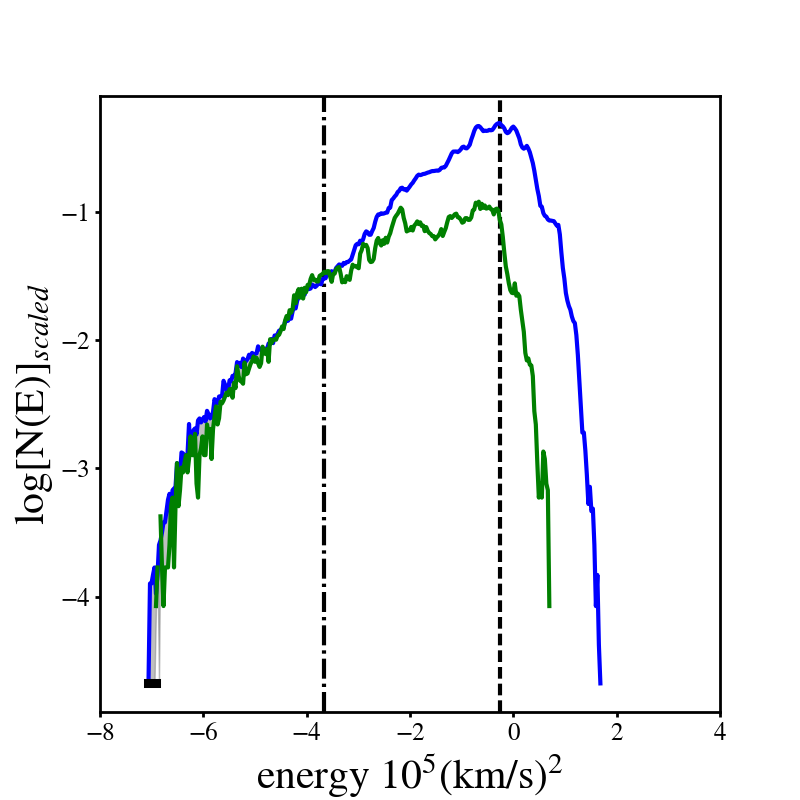}
  \label{fig:sub2}
\end{subfigure}
\caption{An illustration of how we quantify differences between two $N(E)$s.  The energy distributions for Set A (green) and Set A$^c$ (blue) 
at $z=1.47$ (left) and $z=0$ (right) after normalization.  The dashed line indicates the 
energy associated with the peak of $N(E)$ for Set A$^c$, and the dashed-dotted line indicates the midpoint between the dashed line and the most bound energy 
of the system.  All particles with energies less than this value are used in the calculation of the residual, given in equation \ref{diffcalc}, and shown as the shaded region 
between the two curves.  The horizontal black line segment shows the difference in energy between the most bound Set A and Set A$^c$ particle; its evolution is plotted 
later in Figure \ref{avebounddiff}.}
\label{NEprocedure}
\end{figure*}

We are also interested in assessing how mixed the particles are in energy.  We compare the shape of the energy distribution of 
the Set A particles with that of all other halo dark matter particles, Set A$^c$, at the most bound energies.  To accomplish 
this, we define an energy range over which we make the comparison.  We take only particles within $r_{vir}$, and then define an energy 
cutoff that equals the energy associated with the peak of $N(E)$ (dashed line in Figure \ref{NEprocedure}). Because we are interested in the most bound energies in the halo, we take only particles that have 
more negative energies than the midpoint between our cutoff energy and the most bound particle energy\footnote{At the least bound 
energies, the halo energy distribution begins to decrease because the edge of the halo is defined by the FoF and 
not whether the particles are gravitationally bound to the halo.} 
(dashed dotted line in Figure \ref{NEprocedure}).  For example, we use all particles with $E\lesssim-3.08\times10^5$ (km/s)$^2$ at $z=1.49$ (Figure \ref{NEprocedure}).  
Although the midpoint is arbitrary, the result is robust with respect to small changes in the energies we use.  Finally, we normalized the distributions 
before calculating the difference as seen in Figure \ref{NEprocedure}.  This procedure is used for all subsequent plots where we calculated the difference between halo particle distributions.  
We calculate the average difference, or the residual, in $\log[N(E)]$ per energy bin, over $n$ bins, defined as 
\begin{equation}\label{diffcalc}
 \dfrac{1}{n}\sum_{i=1}^n[\log N_{Set A}(E_i)-\log N_{Set A^c}(E_i)].
\end{equation}
The residual at each energy is the magnitude of the range in grey at that energy.  We note that Set A$^c$ membership is updated as new particles 
are accreted, and by $z=0$, contains approximately $1.7\times10^6$ dark matter particles within $r_{vir}$, compared to $5.2\times10^5$ particles within 
$r_{vir}$ at $z=2$.   

\begin{figure}
\centering
\includegraphics[scale=.4]{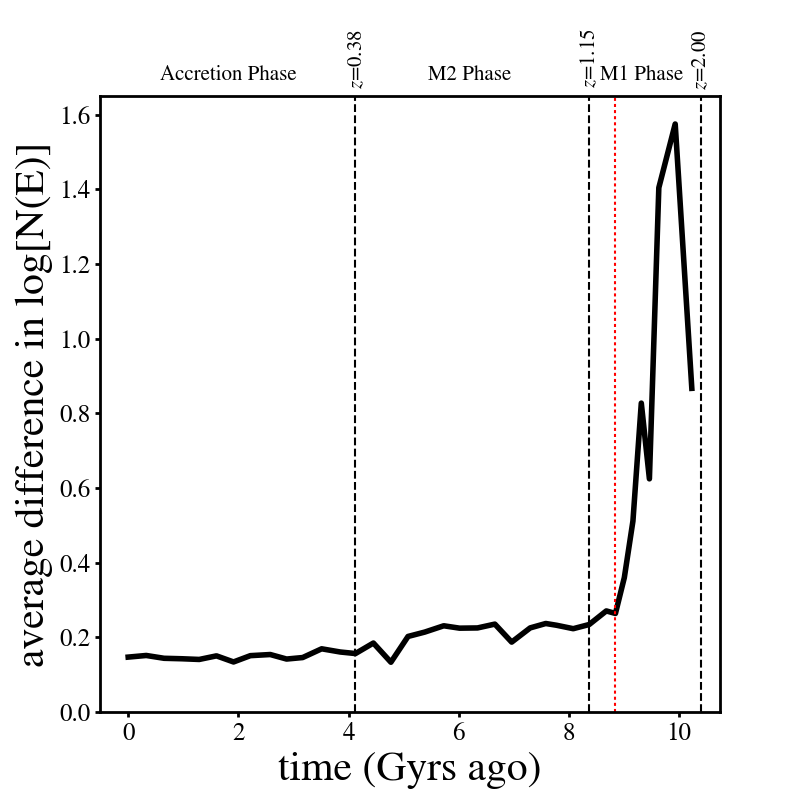}
\caption{The average difference in $\log[N(E)]$ per energy bin of two dark matter particle populations, Set A and Set A$^c$, plotted against cosmic time.  
The black vertical dashed lines indicate the boundaries of the three phases and M1 Phase and M2 Phase stand for Merger 1 and Merger 2, 
respectively.  The red dotted vertical line shows the time when the bump due to the merging halo in $E-L^2$ space, described in Figure \ref{L2vEcontour} 
disappears as the merging halo core is incorporated into the main 
halo.  The curve has a clear downward trend as the two populations of dark matter particles are tending towards the same distribution.  The bend of the curve 
around $\sim$8.4 Gyrs ago, or $z\sim1.3$ coincides with our merger completion criterion, when the centers of mass are within 0.07$r_{vir}$.  The membership of Set A$^c$ 
particles is continually updated as new dark matter particles are accreted over time.}
\label{avediff}
\end{figure}

Figure \ref{avediff} shows this residual across time.  The two distributions are becoming more similar in shape over cosmic time, implying that the 
particles of the two halos are mixing in energy.  The black dashed lines indicate our phase boundaries and the red dashed line 
shows the time when the merging halo's $N(E,L^2)$ is no longer distinguishable as a bump on the $N(E,L^2)$ distribution of the main halo (Figure \ref{L2vEcontour}). The 
bend at around $z\sim1.3$, or approximately 8.4 Gyrs, is a direct result of the core of the smaller halo merging with the center of the main halo.  After 
$z\sim1.3$, the mixing in energy of Set A dark matter particles and the main halo dark matter particles proceed much slower. We should note that a 
significant portion of our average difference per energy bin comes from the most bound energies where the Set A particles have not reached the deepest 
potential yet.  However, the difference between the energy of the most bound particle for the two distributions is decreasing with time and will be 
discussed later in Section \ref{DMmig}.  

Figure \ref{avediff} points toward two timescales for the decrease in the difference of these two $N(E)$s.  There is a rapid change in Merger 1, during the 
initial merger lasting $\sim$1 Gyr, and a slower change over Merger 2 and the Accretion phase lasting $\sim$8.5 Gyrs.  Usually, the dynamical time scale is used as an 
approximation for relaxation time, but exact timescales have not been investigated in the literature.  While we do not measure relaxation times, we use the 
mixing timescale as a proxy since they are both driven by the same physical dynamics.  Here we measure, for the first time, the mixing, and hence, relaxation timescale for a 
halo merger, based on the similarities in energy distribution shape.  While the initial, fast timescale is well approximated by the dynamical timescale, the 
second, slower one is considerably longer. In fact, it is surprising that Set A particles that merged at $z=2$ are still not completely mixed by $z=0$.

\subsubsection{Energy migration of the most bound dark matter particles}\label{DMmig}

\begin{figure}
\centering
\includegraphics[scale=.4]{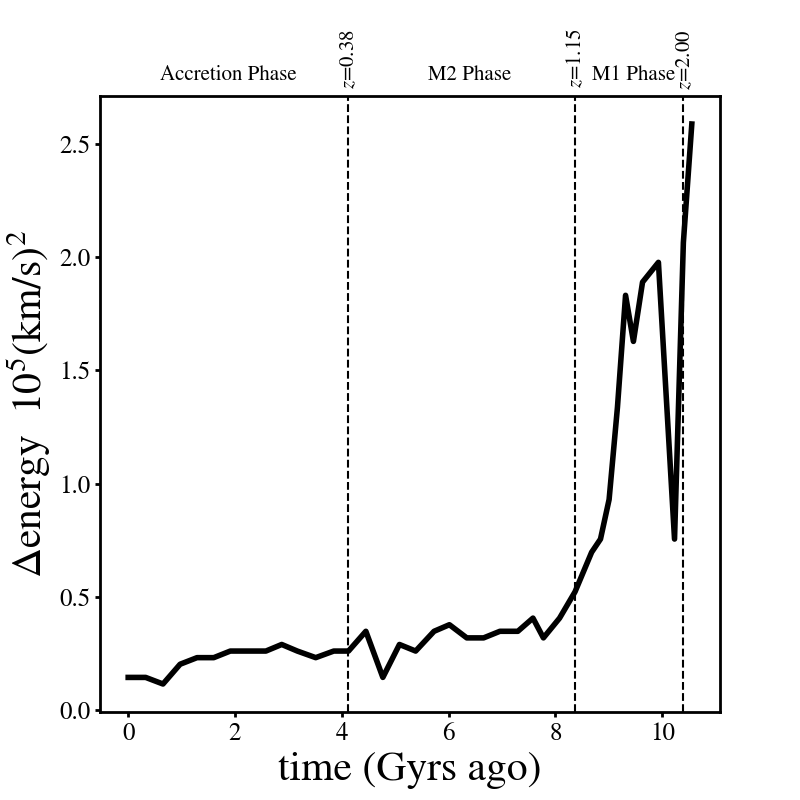}
\caption{The difference between the energy of the most bound dark matter particle of halo 138 and that of the most bound Set A dark matter particle plotted 
against cosmic time.  The black vertical dashed lines indicate the boundaries of the three phases and M1 Phase and M2 Phase stand for Merger 1 and Merger 2, 
respectively.  The difference is decreasing as the most bound Set A particle deepens in potential over time compared to the most bound halo dark matter particle.  
Though similar in shape, what is being plotted here is not the same as in Figures \ref{avediff}, \ref{dmave_diff}, and \ref{allmatterave_diff} where we show a difference in energy 
distributions averaged over many energies.}
\label{avebounddiff}
\end{figure}

The average dark matter particle is losing energy, i.e., falling into a deepening potential as more material is being accreted.  This is true for Set A 
particles as well, and is supported by negative median values for $f_E$ in Figure \ref{dEL2}.  However, as more material is 
being accreted, the most bound of Set A particles fall more rapidly into the potential well than the most bound Set A$^c$ particles.  This can be seen in 
Figure \ref{avebounddiff} where the difference in energy between the halo's most bound dark matter particle and that of the most bound Set A dark matter particle is plotted 
against cosmic time.  The difference decreases towards current time.
The rate of the decrease is not constant, but mimics that seen in Figure \ref{avediff}.  Note, however, that the two Figures plot different quantities, so 
the fact that the two have similar appearances lends additional support for our claim of a fast mixing timescale, followed by a slower one.  Further evidence 
is provided by Figure \ref{dEL2}, which shows that the Set A particles on average lose the greatest fraction of their 
energy during Merger 1, which corresponds to the fast mixing timescale, and then Merger 2, and have the smallest fractional change during Accretion phase.   

Particles falling deeper into the potential and losing energy should have a corresponding decrease in radius.  However, we are unable to disentangle this 
motion from motion in the normal orbit of the particle, say from apocenter to pericenter, in the time resolution of Illustris.

\subsection{Dark matter relaxation}\label{DMrelax}

Having shown that dark matter particles of the main and merging halo are mixing over time, we now would like to determine if this mixing between two halos is leading to 
relaxation.  Because there is no analytical expression for  $N(E,L^2)$ of a relaxed system, we carry out the analysis in energy 
only.  To assess the degree of relaxation, we perform a calculation similar to the one in Section 2.2.2, but in this case we compare the $N(E)$ of all dark matter particles 
to DARKexp $N(E)$, a fully relaxed distribution.  We note that previous works have fitted simulated and observed $N(E)$ distributions with 
DARKexp \citep{Ber13,Hjo15,Ume16,Nol16,You16}. Our emphasis in this paper is not on the fits, but on the residuals between the energy distributions of the 
Illustris halos and DARKexp, and their time evolution.

\begin{figure}
\centering
\includegraphics[scale=.4]{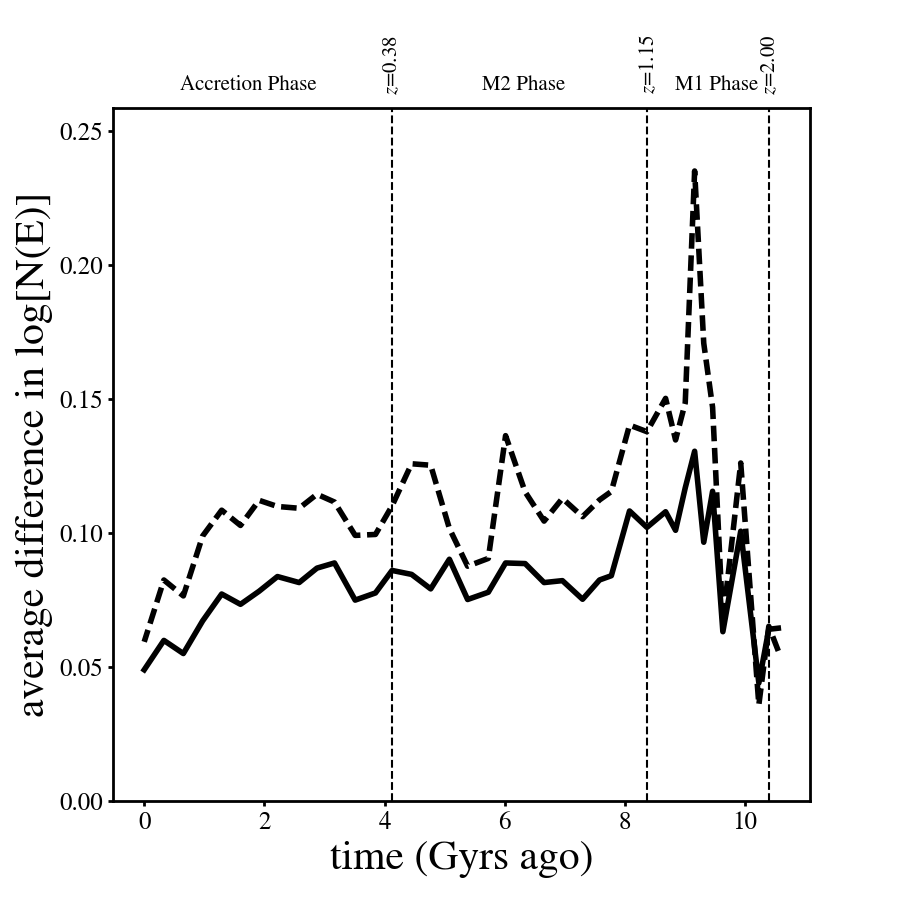}
\caption{The average difference in $\log[N(E)]$ per energy bin of all dark matter particles (the green line plotted in Figure \ref{138NE}) and the DARKexp best 
fit to the dark matter particles, plotted against cosmic time.  The solid (dashed) line includes all particles with energies more bound than 50\% (70\%) 
between the most bound particle and the energy associated with the 
peak of the dark matter energy distribution.  The black vertical dashed lines indicate the boundaries of the three phases and M1 Phase and M2 Phase stand for Merger 1 and Merger 2, 
respectively.  The 
lines show a clear downward trend, especially after the Set A particles complete their 
initial major merger indicated by the Merger 2 - Merger 1 boundary.  The membership of Set A$^c$ is continually updated as new dark matter 
particles are accreted over time by the halo.}
\label{dmave_diff}
\end{figure}

The average difference, or residual, in $\log[N(E)]$ per 
energy bin over cosmic time is shown in Figure \ref{dmave_diff}.  Again, we used the more bound end of the distribution as shown in Figure \ref{avediff} but 
with DARKexp as one of the two distributions.  The 
solid line includes all particles with energies more bound than the midpoint (50\%) between the most bound particle and the energy associated with the 
peak of the dark matter energy distribution; the dashed line includes particles within 70\%, and is shown to indicate the degree of robustness of this measure.  We see a distinct 
downward trend, especially after the Set A particles complete their first major merger, indicating that the dark matter distribution is becoming more like DARKexp 
over time, and therefore more relaxed.  

Note that about 10 Gyrs ago, the difference between the two energy distributions is small.  At this 
time, the initial major merger is underway but most of the merging dark matter particles are outside of our energy cutoff (dashed line in Figure \ref{NEprocedure}).  Only the particles of the central 
main halo are within the cutoff, and because they have a history that leads them to be more relaxed at this time, their $N(E)$ is well approximated by DARKexp.  
As merging halo particles of Set A move to more bound energies, they pass our 
energy cutoff and are included in our calculation (similar to equation 2.2).  This results in the residuals in $N(E)$s increasing to its peak 
value around 8.5 Gyrs ago (solid line in Figure \ref{dmave_diff}).  After the majority of merging particles are included, the halo begins to show signs of relaxing i.e., the difference decreases 
until the second merger event begins.  Just like the prior major merger, the merging halo is not included in the difference calculation at first.  The 
central halo continues to relax and the average difference decreases from $\sim$8.5 Gyrs ago to $\sim$7 Gyrs ago. Throughout the rest of Merger 2, there are 
two competing effects: the central portion of the halo is relaxing, and unrelaxed particles are starting to be included at less bound energies. These two 
effects appear to negate each other causing no bulk average change in the difference until Accretion phase, when no more major mergers occur and the entire halo 
begins to relax.  

Unlike Section \ref{DMmix} that discussed mixing, this is a direct comparison to a theoretical model for relaxed systems and 
therefore describes the relaxation state of the halo.  The downward trend in Figure \ref{dmave_diff} starts at the same time as the rapid change in 
Figure \ref{avediff}, although the shape of the downward trend is different.  The approximate similarity of the shape of the curves in Figures \ref{avediff}, 
\ref{avebounddiff}, and \ref{dmave_diff} supports the notion that mixing between two halos and relaxation are driven by the same dynamics; energy is changed 
by interactions with the global time-varying potential and angular momentum is changed by torques caused by asymmetries.  We note that we use Figure 
\ref{avediff} to measure the timescale because the curve is less noisy than that in Figure \ref{dmave_diff}.

\section{Dynamical evolution of baryons in dark matter halos}\label{DEB}

\begin{figure*}
\centering
\begin{subfigure}{.5\textwidth}
  \centering
\includegraphics[width=\columnwidth]{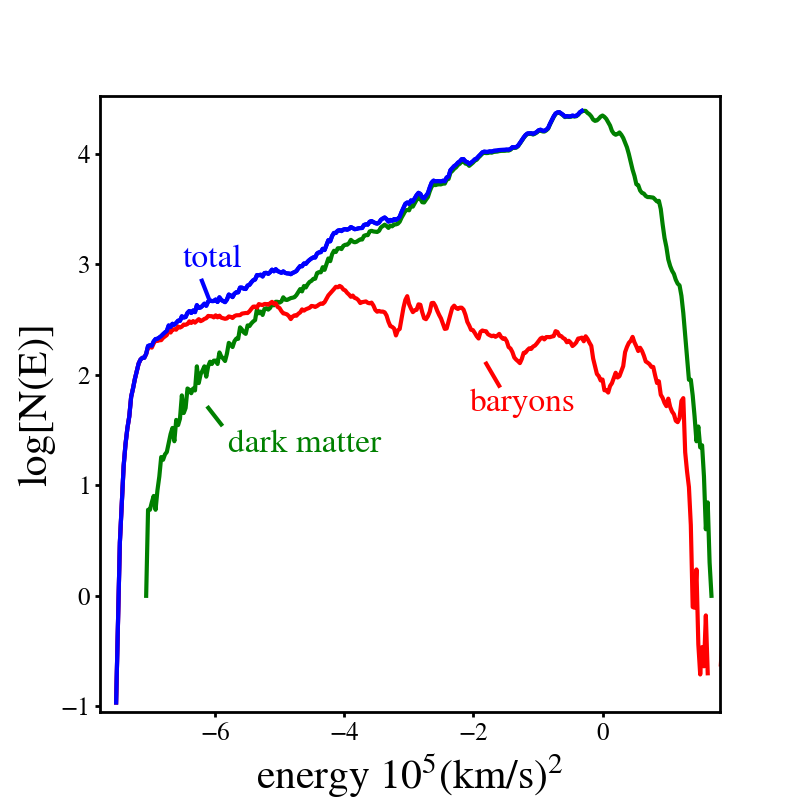}
  \label{fig:sub1}
\end{subfigure}%
\begin{subfigure}{.5\textwidth}
  \centering
\includegraphics[width=\columnwidth]{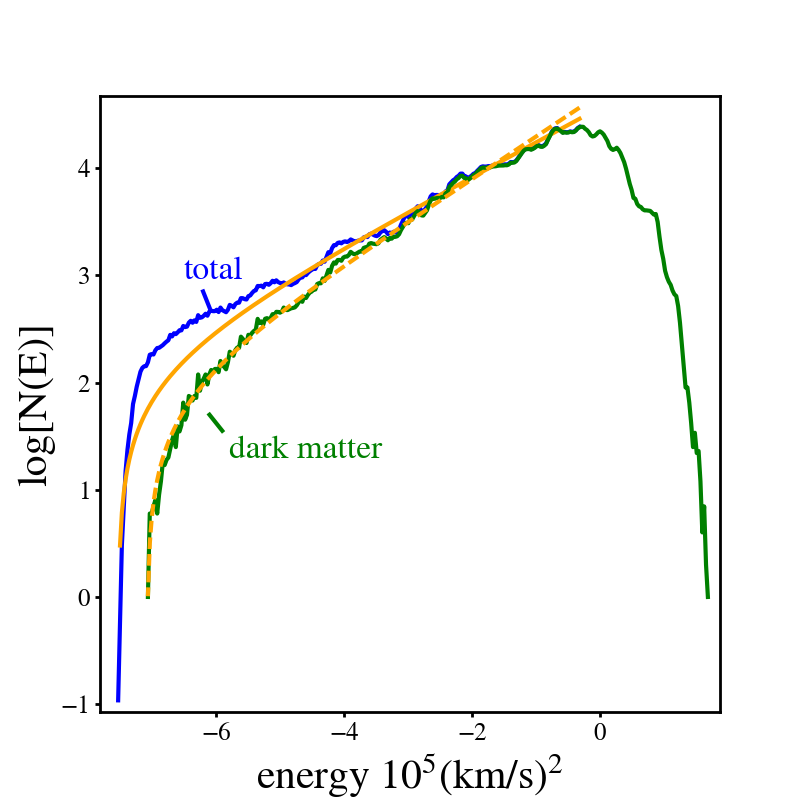}
  \label{fig:sub2}
\end{subfigure}
\caption{Left: The energy distribution of halo 138 at $z=0$.  The blue line is the truncated total matter energy distribution.  The distribution 
labeled `total' includes only particles with energies more bound than the energy associated with the peak of the energy distribution.  The 
other two lines represent the energy distributions of baryons (red) and dark matter (green).  Right: The total matter and dark matter energy distributions 
from the left panel as well as the DARKexp best fit to the total matter energy distribution (orange solid), and best fit to the dark matter energy 
distribution (dashed orange).  The DARKexp profile, dashed line, is on top of the green dark matter profile on the right.  DARKexp does a 
poor job of fitting the total matter distribution at the most bound end (large negative $E$) because that region is dominated by stars, which are far from 
relaxed and have a flatter distribution across energies. 
}
\label{138NE}
\end{figure*}

Having discussed the mixing and relaxing of dark matter particles, we now want to address the evolution of the entire system, including baryons.  We will look 
for the signature of relaxation in a similar way as before, by comparing the total matter $N(E)$ to DARKexp in Section 3.1.  Particle mixing, however, will be described in a 
different way, in Section 3.2.

Illustris has both star and gas particles.  Baryons are handled differently within Illustris, including having a smaller gravitational softening length and smaller 
mass resolution than dark matter.  
However, if we account for the different masses between the dark matter particles and baryon particles, we can create properly weighted energy distributions and 
we combined the two to make a total matter $N(E)$.  

We plot the halo 138 $z=0$ energy distribution in Figure \ref{138NE}, 
showing the baryon (red) and dark matter (green) components, as well as the total matter $N(E)$ (blue) in the left panel.  The right panel shows the best fit DARKexp model for the total matter 
$N(E)$ (solid orange) and the best fit DARKexp model to the dark matter $N(E)$ (dashed orange) along with the dark matter (green) component and total matter 
$N(E)$ (blue) from the left panel.  The very different $N(E)$s for the baryons and dark matter imply that the two populations are not well mixed.  At all 
energies, including the most bound, baryons have a much flatter energy distribution.  Therefore, they dominate the mass in the center regions of Illustris 
halos.  Compared to baryons, the dark matter profile is better described by the shape of DARKexp.  Since the baryon energy distribution is not relaxed for 
halo 138 even at $z=0$, the overall energy distribution at the most bound energies will have significant departures from a fully relaxed state.

\subsection{Relaxation of the dark matter and baryon system}

\begin{figure}
\centering
\includegraphics[scale=.4]{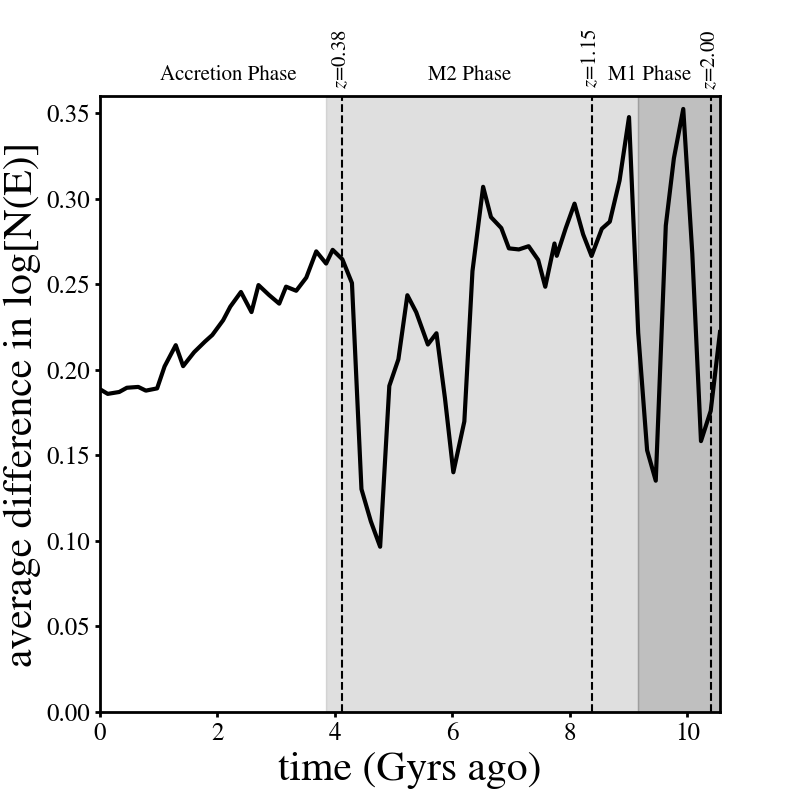}
\caption{The average difference in $\log[N(E)]$ per energy bin of all particles (total matter) in the system and the DARKexp best fit, plotted against cosmic 
time.  The black vertical dashed lines indicate the boundaries of the three phases and M1 Phase and M2 Phase stand for Merger 1 and Merger 2, 
respectively.  After the last major merger, indicated by the leftmost 
dashed vertical line, the difference decreases steadily.  Epochs when $3<N_{star}(E)/N_{gas}(E)<10$ are shaded in light grey, and epochs when 
$N_{star}(E)/N_{gas}(E)<3$ are shaded in dark grey.  During Accretion phase, when the ratio of $N_{star}(E)/N_{gas}(E)>10$, the system contains mainly collisionless 
particles and can be compared to DARKexp.}
\label{allmatterave_diff}
\end{figure}

The mixing shown in Figure \ref{avediff} and the increasing similarity between dark matter energy distribution and DARKexp shown in Figure \ref{dmave_diff} indicate the halo is 
becoming more relaxed.  We now quantify the residual between DARKexp and $N(E)$ of the whole system
in the same way as before.  To find the best fit DARKexp, we again truncated the data to include only particles that are more bound than the energy associated with the peak of the 
energy distribution, as in Figure \ref{NEprocedure}.  To do the fit, we also exclude the most bound particles, as they are dominated by stars which are far 
from relaxed.  If we did include this portion, the fit would be significantly biased by the unrelaxed baryon distribution (see Figure \ref{138NE}), 
rendering the detailed comparison with the relaxed DARKexp model less meaningful. We do, however, use the most bound particle energy value, which is due to 
stars, to constrain our fit. The average difference in $\log[N(E)]$ per energy bin over the time evolution of the system is plotted in 
Figure \ref{allmatterave_diff}.

The interpretation of the difference value is complicated by the complex physics baryons undergo during mergers, such as spatially different distributions of baryons 
and dark matter and increased star formation, which creates new stellar particles.  Because stars are collisionless, while gas is not, the total $N(E)$ distribution can only be modeled with DARKexp if gas 
contributes a negligible amount of particles over the energy range we are considering.  This is often the case for the system in later epochs, after the last 
major merger.  We used a (mass-weighted) ratio $N_{star}(E)/N_{gas}(E)$ over the energies where the average difference was calculated for Figure 
\ref{allmatterave_diff}, to determine when gas was negligible.  
In Figure \ref{allmatterave_diff}, epochs when $3<N_{star}(E)/N_{gas}(E)<10$ are shaded in light grey, and epochs when $N_{star}(E)/N_{gas}(E)<3$ are shaded in dark 
grey.  The system has a factor of 10 more stars than gas during nearly all of the Accretion phase, as most of the gas has been expelled from the most bound 
energies or turned into stars following the last major merger.  During Accretion phase, which is the last 4 Gyrs, the residual in the energy distributions is 
steadily declining as cosmic time approaches the present.  We conclude that the whole system is becoming more relaxed over time, though the shape of $N(E)$ of 
dark matter particles alone is much closer to DARKexp than the shape of $N(E)$ of all matter (right panel in Figure \ref{138NE}).  

\subsection{Lack of mixing of dark matter and baryons}

We saw earlier in Section \ref{DEB} that baryons preferentially occupy the most bound energy
states in the $N(E)$ distribution (Figure \ref{138NE}), and hence are not well mixed with dark matter
particles.  Given what is
known about galaxy formation, it may not seem surprising that stars ended up as the
most bound particles.  Gas, being highly dissipational, collected at the bottom of
the potential wells and formed stars. Most of the stars formed this way in the
small, high-$z$ progenitor halos that eventually merged to form halo 138. What is
somewhat surprising, or at least deserving of attention, is that the stars stayed as
the most bound particles, through mergers and other evolutionary events. In
principle, they could have mixed with dark matter particles, and acquired an energy
distribution more similar to that of dark matter.

To address this, let us compare the behavior of baryons with that of dark matter particles belonging to a merging halo, i.e., Set A particles.  We saw that Set A and
Set A$^c$ are continually mixing, but dark matter and baryons are not. One possible explanation for the difference, we
argue, is that during a merger, baryons at the center of the smaller halo have very negative energies, so that the fluctuating potential of the resulting halo 
imparts only a small fractional energy change to baryons, and hence stars.  Since mixing is achieved through energy exchange, but stars' energies stay roughly the same, stars stay largely unmixed, and 
remain the most bound and centrally concentrated of all particles.  Energies of less bound particles, on the other hand, being smaller in magnitude, are more 
affected by the time fluctuating potential, allowing for more thorough energy exchange.

\begin{figure}
\centering
\includegraphics[scale=.45]{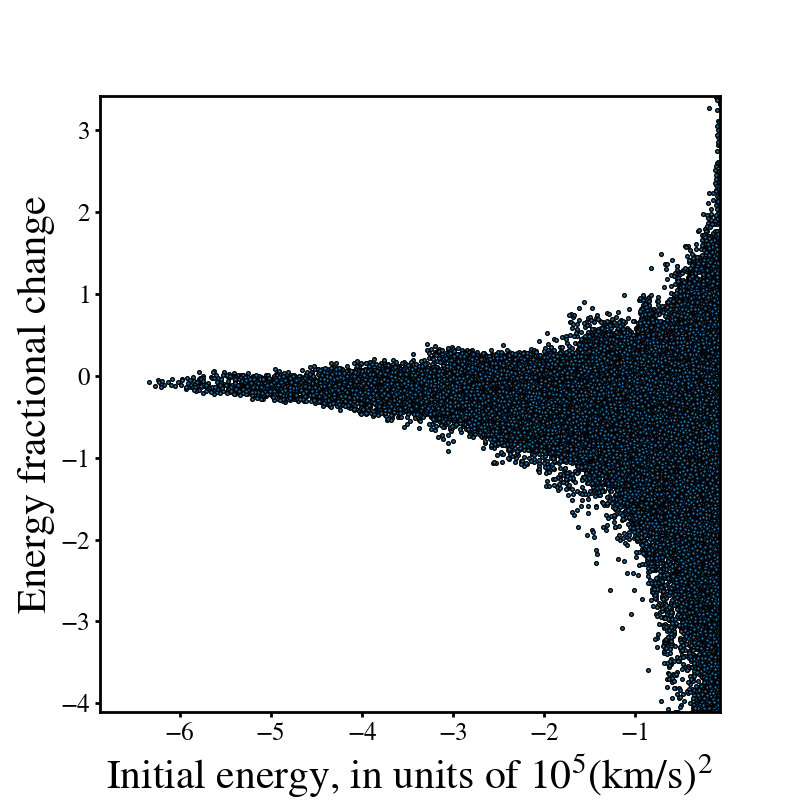}
\caption{The fractional change in energy plotted against the initial energy of Set A particles during Merger 2.  Other phases show similar behavior.  The most bound particles have small $f_E$, 
whereas less bound particles exhibit a greater range of energy redistribution.}
\label{fEvEi}
\end{figure}

This should also apply to more bound dark matter particles.  We can test this by looking at the fractional energy change of Set A  particles as a function of 
their energy.  Figure \ref{fEvEi} confirms that the most bound dark matter particles retain most of their energy. 

One possible mechanism for the difference in fractional energy change, involves the time spent in the fluctuating potential, for different merging halo 
particles.  As the smaller halo starts to 
fall into the main halo, its less bound particles are tidally stripped, resulting in them spending longer times in the outer and intermediate portions of the 
system where they are exposed to the time-fluctuating potential for longer, allowing for better energy exchange.  On the other hand, the core of the merging 
halo, aided by dynamical friction, tends to sink more directly towards the core of the main halo, preserving their very negative total energies. 

\begin{figure}
\hskip -1.5cm
\includegraphics[scale=.45]{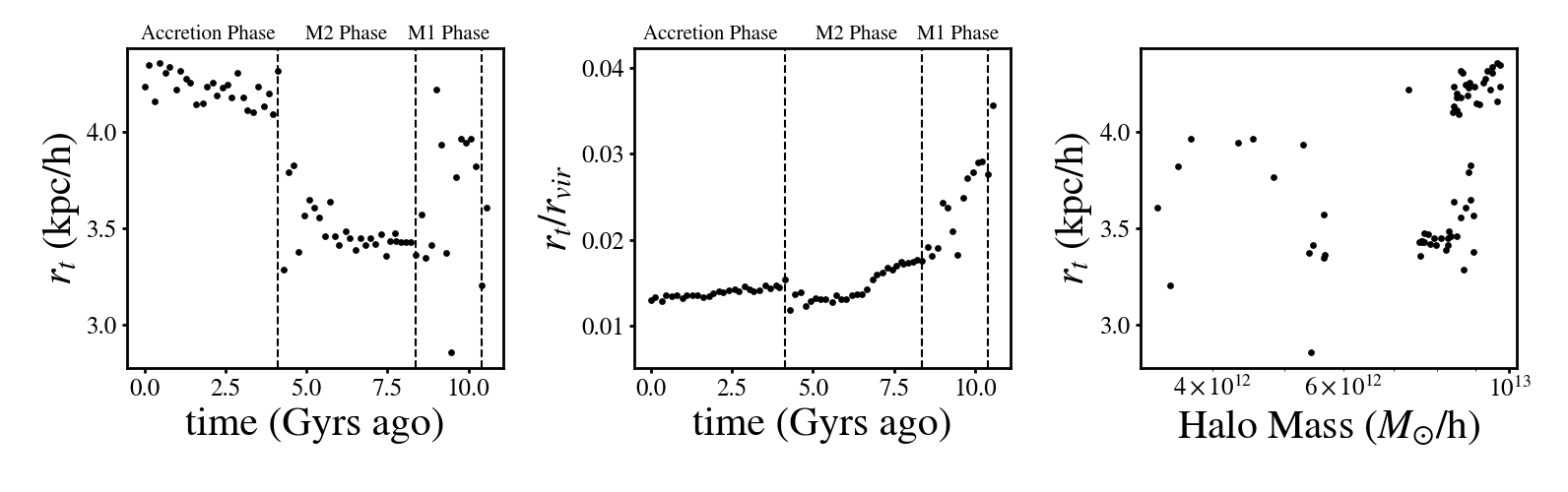}
\caption{The transition radius (left), and transition radius divided by the halo virial radius (middle), plotted against cosmic time for halo 138.  The vertical 
dashed lines are the same as those in Figure \ref{avediff} and indicate the boundaries of the three phases described previously and M1 Phase and M2 Phase stand for Merger 1 and Merger 2, 
respectively.  The right panel is the 
transition radius plotted against halo mass for halo 138.  $r_t$ is increasing as the halo grows but not by as much as $r_{vir}$ is increasing.  This means 
the region of the halo within $r_t$ is becoming a smaller portion of the overall halo volume over time.  In the right panel, it is easy to see the major 
merger events as gaps in the halo mass.}
\label{rt3panel}
\end{figure}

Given enough time, will stars be able to mix more fully with dark matter particles?  The past history of halo 138 may
serve as a guide.  We define a transition radius, $r_t$, as the spherically averaged radius at which the
density of baryons is equal to that of dark matter, and we track this radius through
cosmic time. Figure \ref{rt3panel} shows $r_t$ in physical units (left), and in terms of $r_{vir}$ (middle). The right panel shows the transition radius 
plotted against the halo mass.  The physical size of $r_t$, in kpc, is growing slightly with cosmic time, but not as much as $r_{vir}$, which is increasing 
due to major mergers early in its history, and minor mergers and 
continual mass accretion later in its history.  This leads $r_t/r_{vir}$ to decrease slightly towards the present.  This 
means the region encapsulated by $r_t$ is found at somewhat more bound energies over time.  Because the transition radius has not changed significantly for 
the last 5-8 Gyrs, it is unlikely that it will change for at least several billion years.  This suggests that stars will not mix with the dark matter and 
relax for a very long time.  

\begin{figure*}
\centering
\begin{subfigure}{.5\textwidth}
  \centering
\includegraphics[width=\columnwidth]{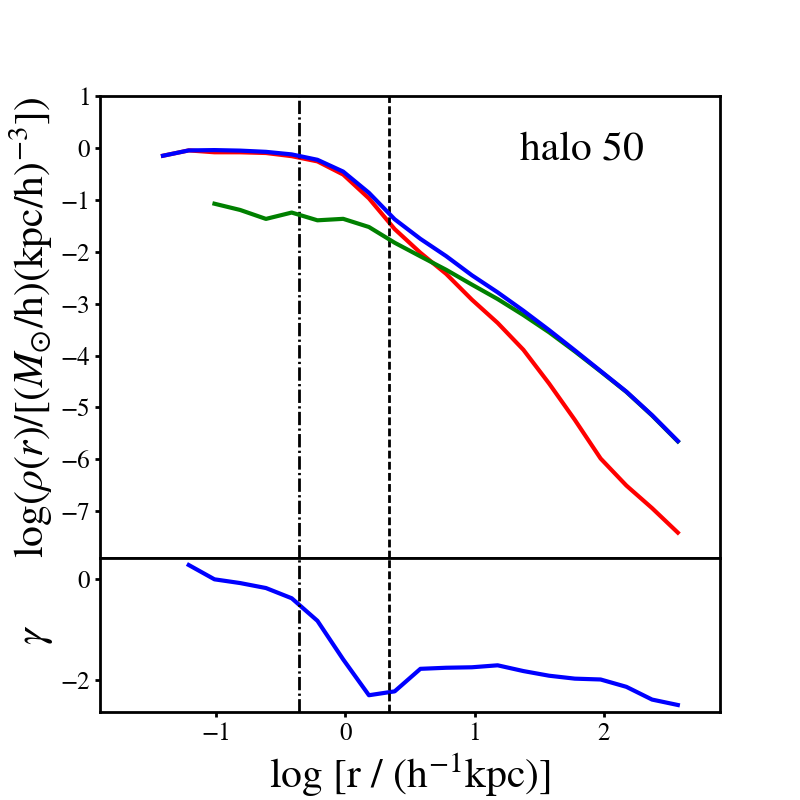}
  \label{fig:sub1}
\end{subfigure}%
\begin{subfigure}{.5\textwidth}
  \centering
\includegraphics[width=\columnwidth]{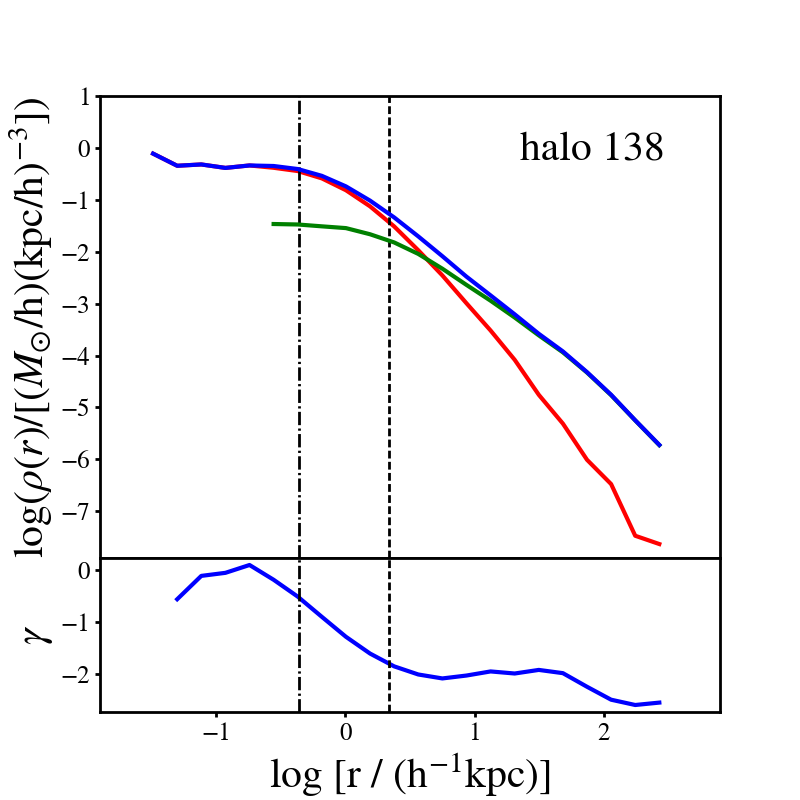}
  \label{fig:sub2}
\end{subfigure}
\caption{Top: The density profile for Illustris-1 halo 50 (left) and halo 138 (right) at $z=0$.  The blue line is the total matter profile while the green is 
dark matter, and the red is baryons. Bottom: The total matter density profile slope, where $\gamma=d\log(\rho)/d\log(r)$. For halo 50, there is a feature at 
$\log[r/(\text{kpc}/h)]\sim 0.25$ in the density profile slope indicating the dark matter to baryon transition region. This feature is not as pronounced in the halo 138 
density profile slope.  The dashed line and the dashed-dot line represent the softening length parameter for dark matter and baryons respectively.}
\label{density}
\end{figure*}

The fact that stars are not well mixed with dark matter, and there exists a well defined transition radius between the two, gives rise to a dimple, or an 
`oscillation', in the density profile slope of observed \citep{Cha14,Tho07} and simulated galaxies \citep{Sch15a, Xu17}.  An 
example of an oscillation can be seen in the density profile and its slope of halo 50 (left) and halo 138 (right), both at $z=0$ (Figure \ref{density}).  The 
blue lines in the top panels show the total matter density while the green and red lines show the dark matter and baryon 
densities, respectively.  The blue line in the bottom panels shows the total matter density slope, defined as $\gamma=d\log(\rho)/d\log(r)$.  The origin of 
these oscillations and their relation to $N(E)$ are examined in Young et al. \citep{You16}. 

\section{Conclusions}\label{Conclusions}

We have studied the dynamical evolution of an Illustris galaxy by tracking the redistribution of particle energy and angular momentum during and after the 
merger process, with the main goal of understanding how a system moves towards a relaxed state.  To our knowledge, this is the first paper that addresses 
relaxation by comparing a galaxy's energy distribution to that of a theoretical prediction for fully relaxed collisionless systems.

We split the halo's merger history into 3 phases, based on key merger events (see Figure 1).  Within each of the phases, we followed the dark matter 
particles from a major merger around $z=2$, which we call Set A particles, and calculated their changes in energies and angular momenta. Across two major 
mergers, we found that dark matter particles generally moved to more bound energies, whereas their change in angular momentum was situation specific 
and related to which population a particle belonged to, merging halo or main halo.  Particles tended to lose $L^2$ when merging with a larger halo, and gain 
$L^2$ when they were a part of the main halo.  In the final phase, which started after the last major merger, the particles showed a near symmetric 
distribution about zero change in $L^2$, indicating that at least some subset of them are well mixed.

Motivated by this observation, we then proceed to look for other, more direct signatures of particle mixing and galaxy relaxation, both in dark matter only, 
and dark matter plus baryon.  In the case of dark matter, we compared the Set A particles' $N(L^2)$ and $N(E)$ distributions to those of the main halo dark 
matter particles with which they were merging.  We found the two populations to be well mixed in $L^2$ for energies that are more bound than $\approx$2/3 the 
halo's most bound energy at $z=0$.  To assess mixing in energy, we calculated the difference in the shape of logarithmic energy distributions of the two 
sets.  We reasoned that given enough time, these two distributions will become the same.  In fact, the difference between the two is steadily decreasing as 
we move towards the present, although at $z=0$ they are still not the same as seen in Figure \ref{avediff}.  

The time evolution can be characterized by two 
timescales. The first describes the epoch during Merger 1 when there is a rapid convergence between the two distribution shapes, as the two halos are 
merging.  It lasts approximately 1 Gyr during which the average difference in the energy distributions drops by $\sim$80\% of its initial value.  During a 
second, longer timescale, the two distributions steadily move towards each other, but at a much slower pace, lasting from the end of the initial major 
merger to present, or about 8.5 Gyrs. During this time the average difference in the energy distributions drops by $\sim$50\%.  The 
boundary between these timescales is the completion of the initial major merger of the system. While the dynamical timescale is often used to represent the 
time frame of relaxation, we have directly measured the mixing time by comparing the shapes of dark matter energy distributions of the two merging systems 
(Figure \ref{avediff}).  

To gauge the degree of relaxation, we compared halo's energy distribution to a theoretical model for relaxed collisionless systems, 
called DARKexp.  We found that the difference between the two distributions decreases after the initial major merger, implying that the dark matter particle 
population is becoming more relaxed over time (Figure \ref{dmave_diff}).  The timescale is similar to that found for mixing, and consistent with the fact that 
mixing and relaxation are driven by the same dynamics: interactions with the global time-varying potential for energy, and torques caused by asymmetries for 
angular momentum.

Next, we considered relaxation and mixing of the whole system, consisting of dark matter and baryons.
First, we did a similar analysis as before.  We compared the energy distribution of total matter to DARKexp, and found that after the last 
major merger, there is a marked trend in time of decreasing average difference between the shapes of the two distributions (Figure \ref{allmatterave_diff}).  
This indicates that the halo is relaxing, when contrasted with its initial post-merger configuration.   

Second, we have investigated the mixing of baryons with dark matter.  From the energy distributions of dark matter and baryons as well as their spatial 
distributions, it is apparent that baryons are found at deeper potentials.  This may not be surprising given our knowledge of galaxy formation, with gas 
cooling and sinking to the center, where stars are then formed.  One might expect dark matter and baryons to mix and relax, as was the case with Set A 
particles and the main halo.  However they have not mixed, as seen in Figure \ref{138NE}.  The reason for this lack of mixing between dark matter and baryons 
(and between the most bound and less bound dark matter particles) appears to stem from particles' varying degrees of ability to exchange energy with the halo.  
There is a difference in the fractional change in energy experienced by the most and least bound particles.  During violent, collisionless relaxation, only 
a small fractional change in energy is imparted to the most bound particles, whereas less bound particles achieve greater fractional changes in energy.  We 
propose that this is in part due to the length of time a particle is exposed to the global time-varying potential.  During a merger, the most bound particles 
of the merging halo tend to fall quickly towards the center, aided by dynamical friction, whereas less bound particles get tidally stripped and spend longer 
times at mid to large radii, and therefore are exposed to the fluctuating global potential for longer, allowing for more energy redistribution.  

This is especially 
pronounced with stars, as they are preferentially found at the most bound energies when compared to dark matter.  This distinction has implications 
for the distribution of matter in the central regions of galaxies, where it is manifested as an oscillation in the density profile slope, marking the transition between the dark matter dominated 
larger radii and the baryon dominated smaller radii. 

Finally, because the baryon-dark matter transition radius has not changed much for the last 8 Gyrs, we speculate that baryons and dark matter will not mix 
for a long time in the future.

\acknowledgments
We would like to thank Dylan Nelson and Mark Vogelsberger for providing clarification of Illustris.

\bibliographystyle{JHEP}
\bibliography{bib}

\end{document}